%% file: PaperMelting.tex
\documentclass[aps, pra, amsmath, amssymb, reprint]{revtex4-1}

\usepackage[pdftex]{graphicx}
\usepackage[pdftex]{hyperref}
\usepackage{bm}

\begin{document}
\title{Stabilization of vortex-liquid state by strong pairing interaction}
\author{Kyosuke Adachi}
\author{Ryusuke Ikeda}
\affiliation{Department of Physics, Kyoto University, Kyoto 606-8502, Japan}
\date{\today}
\begin{abstract}
We theoretically investigate qualitative features of the field-temperature ($H$-$T$) phase diagram of superconductors with strong attractive interaction lying in the BCS-BEC crossover regime. Starting with a simple attractive Hubbard model, we estimate three kinds of characteristic fields, i.e., the pair-formation field $H^*$, the vortex-liquid-formation field $H_\mathrm{c2}$, and the vortex-lattice-formation field $H_\mathrm{melt}$. The region between $H_\mathrm{c2}$ and $H_\mathrm{melt}$, as well as that between $H_\mathrm{c2}$ and $H^*$, is found to be enlarged as the interaction is stronger. In other words, a strong attractive interaction can stabilize both the vortex-liquid and preformed-pair regions. We also point out the expected particle-density dependence of the $H$-$T$ phase diagram.
\end{abstract}
\maketitle

\section{Introduction}

So far, the superfluid transition with variable attractive interaction between Fermions has been studied primarily in the field of the ultracold atomic physics. Physical properties have been investigated especially in the BCS-BEC crossover regime, where the interaction between particles is strong enough to create non-condensed preformed pairs~\cite{Randeria_Taylor_2014}.

Intriguingly, recent experiments have suggested that a strong attractive interaction can exist in FeSe and related superconductors~\cite{Kasahara_Watashige_2014, Rinott_2017}, which can pave the way for material realization of the BCS-BEC crossover. In contrast to the electrically neutral ultracold atoms, electrons in a superconductor are charged and thus naturally coupled with the gauge field of an external magnetic field. Therefore, FeSe and related materials can provide an opportunity to experimentally scrutinize unexplored effects of the magnetic gauge coupling on superconductors with strong attractive interaction. In fact, superconducting-fluctuation effects on diamagnetic response observed in FeSe are unusually enhanced compared with those in conventional superconductors~\cite{Kasahara_Yamashita_2016}, which may be understood as caused by the strong attractive interaction~\cite{Adachi_Ikeda_2017}. In addition, recent NMR measurements have proposed that a pseudogap caused by the preformed-pair formation can exist, and that the onset temperature of the pseudogap depends on the magnetic-field strength~\cite{Shi_Arai_2017}.
The pair-formation field $H^* (T)$ in FeSe has been estimated based on this NMR result~\cite{Shi_Arai_2017} and also on thermodynamic and transport properties~\cite{Kasahara_Yamashita_2016}.
However, theoretical understanding of these magnetic-field effects on superconductors with strong attractive interaction remains incomplete.

The field v.s. temperature ($H$-$T$) phase diagram of a superconductor with strong fluctuation has been thoroughly investigated in relation to high $T_\mathrm{c}$ cuprates~\cite{Fisher_Fisher_1991, Ikeda_1995, Bennemann_2008} which are believed to belong to superconductors with high particle density.
There, it has been clarified by developing the superconducting fluctuation theory~\cite{Fisher_Fisher_1991, IkedaLT_1990} that the so-called upper critical field $H_\mathrm{c2}(T)$ in the three-dimensional (3D) type-II superconductor is not a phase transition line but a crossover one separating the vortex-liquid region, which is the strongly fluctuating region of the normal phase, from the conventional normal phase with negligibly weak fluctuations, and that, in clean 3D materials, the genuine superconducting ordering occurs as a weak first-order transition corresponding to the vortex-lattice melting~\cite{Brezin_Nelson_1985}.
The vortex-lattice melting curve $H_\mathrm{melt} (T)$ can alternatively be determined by examining the elastic energy of the mean-field vortex-lattice state and invoking the Lindemann criterion~\cite{Bennemann_2008}. In the so-called lowest-Landau-level (LLL) approach to the GL theory, it is believed that $H_\mathrm{melt} (T)$ should be found as a consequence of the superconducting fluctuation. In fact, the fluctuation effect shows the scaling behavior of the form $T-T_\mathrm{c}(H) \sim (TH)^{2/3}$~\cite{IkedaLT_1990}, while the field dependence of the melting temperature also obeys this scaling behavior~\cite{Moore_1989}.
Experimentally, the first-order melting transition and the transition line $H_\mathrm{melt} (T)$ have been observed in high $T_\mathrm{c}$ cuprates~\cite{Safar_1992, Zeldov_1995, Welp_1996}.

In this study, we theoretically investigate qualitative features of the $H$-$T$ phase diagram of superconductors with strong attractive interaction. To obtain a qualitative picture, we start with a simple attractive Hubbard model. Using the T-matrix approximation combined with analysis of the Ginzburg-Landau (GL) functional, we estimate three types of characteristic magnetic fields: the pair-formation field $H^*$, the vortex-liquid-formation field $H_\mathrm{c2}$, and the vortex-lattice-formation field $H_\mathrm{melt}$. The region between $H_\mathrm{c2}$ and $H_\mathrm{melt}$, as well as that between $H^*$ and $H_\mathrm{melt}$, is found to become broader as the attractive interaction gets stronger. Based on this result, we conclude that a strong attractive interaction can stabilize both the vortex-liquid and the preformed-pair regions.

\section{Model}
\label{sec:Model}

To consider qualitative magnetic-field effects on electron systems with strong attractive interaction, we begin with a simple attractive Hubbard model on a simple cubic lattice:
\begin{equation}
H = - t \sum_{\langle i, j \rangle, \sigma} \left( c_{i \sigma}^\dag c_{j \sigma} + c_{j \sigma}^\dag c_{i \sigma} \right) - U \sum_i c_{i \uparrow}^\dag c_{i \downarrow}^\dag c_{i \downarrow} c_{i \uparrow}.
\label{eq:Hamiltonian}
\end{equation}
Here, $c_{i \sigma}^{(\dag)}$ is the annihilation (creation) operator of an electron with spin $\sigma$ at site $i$, and $\langle i,j \rangle$ means a nearest-neighbor pair of sites. There are two parameters in our model: the nearest-neighbor hopping amplitude $t (> 0)$ and the onsite attractive interaction $U (> 0)$. For simplicity, the magnetic-field term is introduced at the stage of analyzing our GL functional (see Sec.~\ref{sec:ThreeKindsOfFields} and Appendix~\ref{app:LLL}). Basically, this simplification, equivalent to the electronic semi-classical approximation, corresponds to neglecting the Landau quantization of electron kinetic energy. In the following, the lattice constant is set to unity.

\section{Zero-field pair-formation and pair-condensation temperatures}
\label{sec:ZeroField}

As a preliminary step to explore magnetic-field effects, we estimate the zero-field pair-formation and pair-condensation temperatures. Though the results presented in this section is well-known~\cite{Micnas_1990}, we show them for completeness. As shown in the following, the pair-formation temperature $T^*$ is calculated within the mean-field approximation~\cite{Melo_Randeria_1993, Iskin_Melo_2009}, and the pair-condensation temperature $T_\mathrm{c}$ is calculated within the T-matrix approximation~\cite{Chen_Stajic_2005, Yanase_Yamada_1999, Maly_Janko_1999, Tsuchiya_Watanabe_2009}. The T-matrix approximation can take into account the shift of chemical potential due to superconducting fluctuation, which is important when the attractive interaction is strong, and in addtion the particle density is not so high~\cite{Nozieres_Schmitt-Rink_1985}.

To calculate $T^*$, we apply to Eq.~\eqref{eq:Hamiltonian} the mean-field approximation, or equivalently, combine the following two equations with each other: the condition for divergence of the uniform superconducting susceptibility [see Eq.~\eqref{eq:SCSusceptibility} for its definition]
\begin{equation}
\chi_{\bm{0}}^\mathrm{(SC)} = \infty,
\label{eq:UniformSusceptibilityDivergence}
\end{equation}
and the particle-number conservation for non-interacting particles
\begin{equation}
n = \frac{2}{M} \sum_{\bm{k}} \frac{1}{\exp [(\epsilon_{\bm{k}} - \mu) / T] + 1}.
\label{eq:NumberEqFree}
\end{equation}
Here, we define several symbols: particle density (per site) $n$, chemical potential $\mu$, temperature $T$, the total number of lattice sites $M = M_x M_y M_z$~\footnote{In our numerical calculation, the number of lattice sites is set as $M_x = M_y = M_z = 32$ and thus $M = 32768$.}, the lattice momentum with the periodic boundary condition $k_\alpha = 2 \pi n_\alpha / M_\alpha$ ($- M_\alpha / 2 \leq n_\alpha < M_\alpha / 2$ with $n_\alpha \in \mathbb{Z}$), and the energy dispersion of non-interacting particles $\epsilon_{\bm{k}} = - 2 t (\cos k_x + \cos k_y + \cos k_z)$. The superconducting 
susceptibility with pair (or center-of-mass) momentum $\bm{q}$ is defined as
\begin{equation}
\chi_{\bm{q}}^\mathrm{(SC)} = \frac{\chi_{\bm{q}}^{(0)} (0)}{1 - U \chi_{\bm{q}}^{(0)} (0)},
\label{eq:SCSusceptibility}
\end{equation}
where
\begin{equation}
\chi_{\bm{q}}^{(0)} (\mathrm{i} \omega_m) = \frac{T}{M} \sum_{\bm{k}, n} G_{\bm{k} + \bm{q}}^{(0)} (\mathrm{i} \varepsilon_n + \mathrm{i} \omega_m) G_{- \bm{k}}^{(0)} (- \mathrm{i} \varepsilon_n).
\end{equation}
Here, $\varepsilon_n = 2 \pi (n + 1 / 2) T$ ($\omega_m = 2 \pi m T$) is the Fermion (Boson) Matsubara frequency, and $G_{\bm{k}}^{(0)} (\mathrm{i} \varepsilon_n) = (\mathrm{i} \varepsilon_n - \epsilon_{\bm{k}} + \mu)^{-1}$ is the non-interacting Green's function.

As for $T_\mathrm{c}$, we apply the T-matrix approximation. This approximation combines the divergence of the susceptibility $\chi_{\bm{0}}^\mathrm{(SC)} = \infty$, which is the same condition as defining $T^*$, with the particle-number conservation
\begin{equation}
n = \frac{2 T}{M} \sum_{\bm{k}, n} G_{\bm{k}} (\mathrm{i} \varepsilon_n) \mathrm{e}^{+ \mathrm{i} \varepsilon_n 0},
\end{equation}
in which superconducting-fluctuation effects are taken into account. Here, $G_{\bm{k}} (\mathrm{i} \varepsilon_n)$ is the interacting Green's function, which is defined as 
\begin{equation}
G_{\bm{k}} (\mathrm{i} \varepsilon_n)^{-1} = G_{\bm{k}}^{(0)} (\mathrm{i} \varepsilon_n)^{-1} - \Sigma_{\bm{k}} (\mathrm{i} \varepsilon_n),
\end{equation}
and $\Sigma_{\bm{k}} (\mathrm{i} \varepsilon_n)$ is the self energy defined within the T-matrix approximation as
\begin{align}
\Sigma_{\bm{k}} (\mathrm{i} \varepsilon_n) = - \frac{T}{M} \sum_{\bm{q}, m} & G_{\bm{q} - \bm{k}}^{(0)} (\mathrm{i} \omega_m - \mathrm{i} \varepsilon_n) \nonumber \\
& \times \frac{U^2 \chi_{\bm{q}}^{(0)} (\mathrm{i} \omega_m)}{1 - U \chi_{\bm{q}}^{(0)} (\mathrm{i} \omega_m)} \mathrm{e}^{+ \mathrm{i} (\omega_m - \varepsilon_n) 0}.
\end{align}
Here, the temperature-independent Hartree shift
\begin{equation}
\Sigma^\mathrm{(H)} = - U \frac{T}{M} \sum_{\bm{k}, n} G_{\bm{k}} (\mathrm{i} \varepsilon_n) \mathrm{e}^{+\mathrm{i} \varepsilon_n 0} = - \frac{U n}{2},
\end{equation}
is already taken into account by properly choosing the origin of energy; therefore we do not explicitly consider $\Sigma^\mathrm{(H)}$~\cite{Yanase_Yamada_1999, Adachi_Ikeda_2018}.

To explain physical meanings of the definitions of $T^*$ and $T_\mathrm{c}$, it is convenient to consider the opposite limit to the weak-coupling BCS one in which $T^*$ and $T_\mathrm{c}$ take almost the same value.
In this strong-coupling limit ($U / t \rightarrow \infty$), we can show that $T^* \propto |\mu| \propto U \propto E_\mathrm{b}$, where $E_b$ is the two-particle binding energy~\cite{Adachi_Ikeda_2018}; therefore, $T^*$ can be interpreted as the pair-formation (or pair-breaking) temperature.
Actually, regardless of the interaction strength $U$, $T^*$ is of the same order of magnitude as the zero-temperature excitation energy gap $E_\mathrm{gap}$, which can be interpreted as a typical energy scale to break an electron pair (see Appendix~\ref{app:Delta}); thus, $T^*$ can be interpreted as the pair-formation temperature.
As for $T_\mathrm{c}$, in the strong-coupling limit, we obtain an asymptotic formula $T_c \propto t^2 / U$, which represents the BEC transition temperature of non-interacting Bosons (or preformed-pairs) with a nearest-neighbor hopping amplitude $t_\mathrm{B} \propto t^2 / U$~\cite{Micnas_1990}; accordingly, $T_\mathrm{c}$ can be understood as the pair-condensation temperature.

\begin{figure}[tbp]
\includegraphics[scale=0.7]{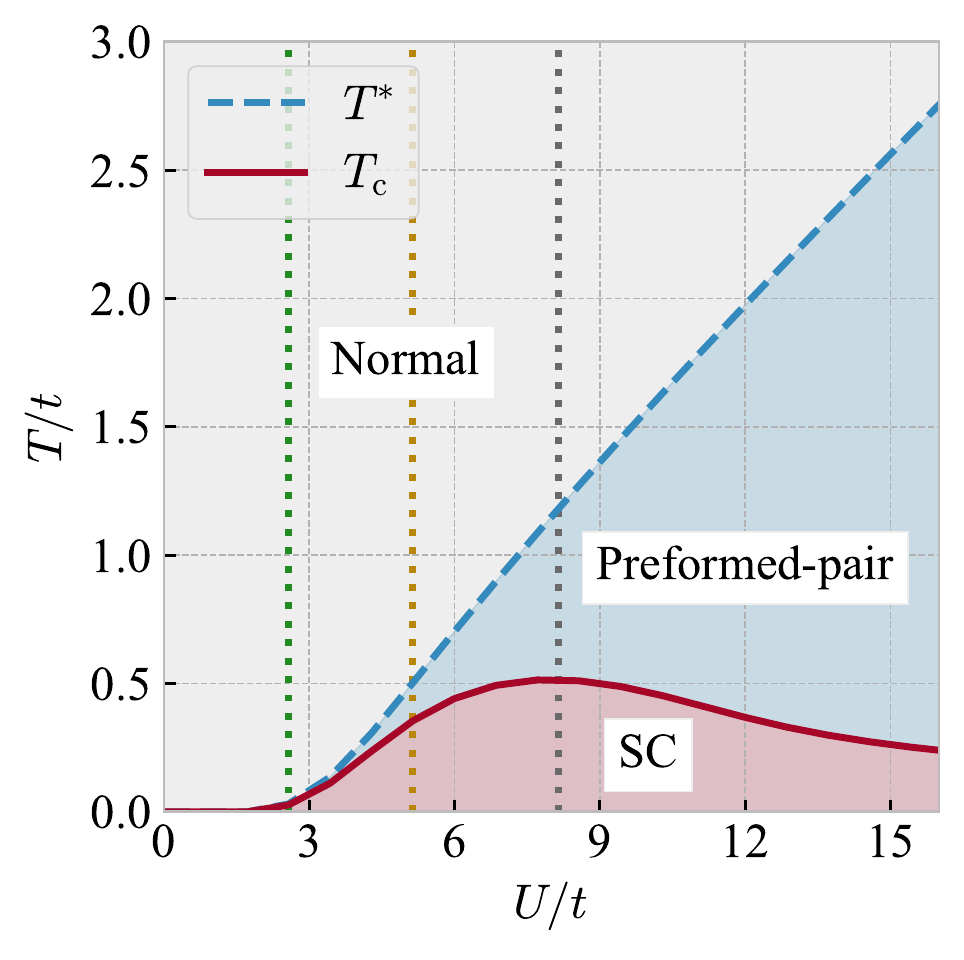}
\caption{Interaction strength v.s. temperature phase diagram for the particle density $n = 0.2$ in zero field. The pair-formation temperature $T^*$ (blue dashed line) roughly separates the normal-state region (grey area) from the preformed-pair region (blue area). The pair-condensation, or superconducting transition, temperature $T_\mathrm{c}$ (red solid line) separates the preformed-pair region from the superconducting (SC) region (red area). The threshold interaction value $U = U_0 \simeq 8.14 t$ for the formation of a two-particle bound state (grey dotted line) and the values of interaction used in the analysis in Sec.~\ref{sec:FieldTemperaturePhaseDiagram}, $U = 2.57 t$ (green dotted line) and $5.14 t$ (yellow dotted line), are also shown.}
\label{fig:TstarTc-U}
\end{figure}

Figure~\ref{fig:TstarTc-U} shows an interaction strength v.s. temperature phase diagram obtained from the equations listed above with the particle density fixed to $n = 0.2$. As seen from Fig.~\ref{fig:TstarTc-U}, the preformed-pair region becomes broader as the interaction gets stronger. In Fig.~\ref{fig:TstarTc-U}, we also show with a grey dotted line the threshold value $U = U_0 \simeq 8.14 t$ for the formation of a two-particle bound state~\cite{Chen_Kosztin_1999, Maly_Janko_1999, Adachi_Ikeda_2018}. Note that the BCS-BEC crossover occurs close to $U_0$.

\begin{figure}[tbp]
\includegraphics[scale=0.7]{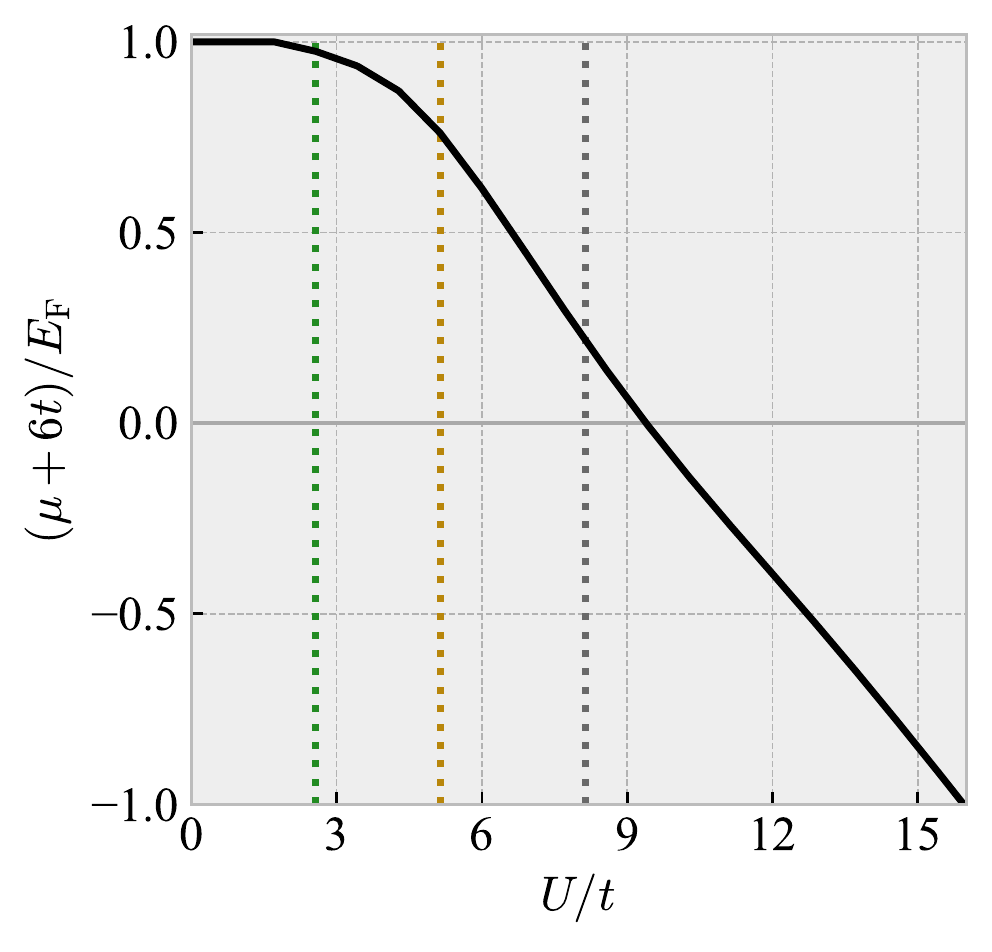}
\caption{Chemical potential $\mu$ at the pair-condensation temperature $T_\mathrm{c}$ for the particle density $n = 0.2$. The vertical axis is measured from the bottom of the non-interacting energy band $-6 t$ in units of the Fermi energy $E_\mathrm{F}$ ($\simeq 2.8 t$). In the same way as Fig.~\ref{fig:TstarTc-U}, the threshold value of interaction $U = U_0 \simeq 8.14 t$ for the formation of a two-particle bound state (grey dotted line) and the values of interaction used in the analysis in Sec.~\ref{sec:FieldTemperaturePhaseDiagram}, $U = 2.57 t$ (green dotted line) and $5.14 t$ (yellow dotted line), are also shown.}
\label{fig:muc-U}
\end{figure}

As shown in Fig.~\ref{fig:muc-U}, the chemical potential $\mu$ is remarkably reduced when the attractive interaction $U$ approaches $U_0$. When $U$ is larger than $U_0$, $\mu$ tends to become lower than the band bottom.

In the following, we focus on systems where $U < U_0$ is satisfied so that the decrease in $\mu$ is not so large. More specifically, we consider two systems with different values of $U$: $U / t = 2.57$ and $U / t = 5.14$ (see the green and yellow dotted lines in Figs.~\ref{fig:TstarTc-U} and \ref{fig:muc-U}).

\begin{figure*}[tbp]
\includegraphics[scale=0.65]{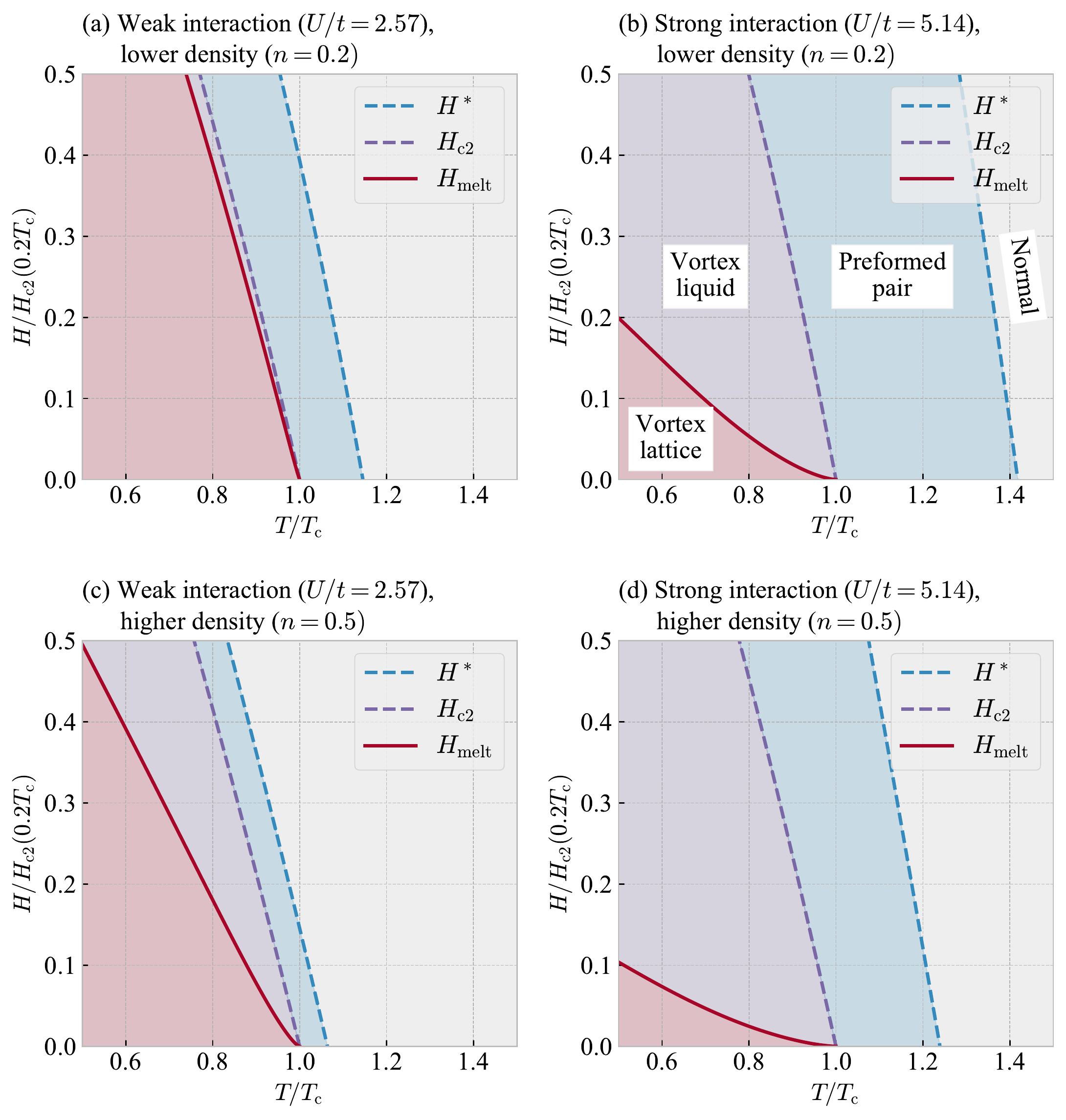}
\caption{Theoretical $H$-$T$ phase diagrams. Figures (a) and (b) respectively show the weak-interaction ($U / t = 2.57$) and strong-interaction ($U / t = 5.14$) cases with lower density ($n = 0.2$). Figures (c) and (d) respectively show the weak-interaction ($U / t = 2.57$) and strong-interaction ($U / t = 5.14$) cases with higher density ($n = 0.5$). In each figure, the pair-formation field $H^*$ (blue dashed line) separates the normal-state region (grey area) from the preformed-pair region (blue area), the vortex-liquid-formation field $H_\mathrm{c2}$ (purple dashed line) separates the preformed-pair region from the vortex-liquid region (purple area), and the vortex-lattice-formation field $H_\mathrm{melt}$ (red solid line) separates the vortex-liquid region from the vortex-lattice region (red area). In all data, the phenomenological parameter to describe $H_\mathrm{melt}$ is fixed as $c = 0.5$ (see the main text).}
\label{fig:H-T}
\end{figure*}

\section{Pair-formation, vortex-liquid-formation, and vortex-lattice-formation field}
\label{sec:ThreeKindsOfFields}

To understand qualitative features of the $H$-$T$ phase diagram, we estimate three kinds of magnetic-field values: the pair-formation field $H^*$, the vortex-liquid-formation field $H_\mathrm{c2}$, and the vortex-lattice-formation field $H_\mathrm{melt}$. In the following, the direction of magnetic field is fixed in parallel to the $z$ axis, and we assume strongly type-II systems and neglect the difference between the applied magnetic field and the magnetic field in the system ($\bm{B} = \mu_0 \bm{H}$). As mentioned in Sec.~\ref{sec:Model}, we neglect the Landau quantization of the electron kinetic energy.

The pair-formation field $H^*$ is calculated in a similar way to the calculation of $T^*$. To introduce the effect of magnetic field $H$, we only have to replace the condition for divergence of the uniform superconducting susceptibility [Eq.~\eqref{eq:UniformSusceptibilityDivergence}] with that for divergence of a finite-momentum superconducting susceptibility~\cite{Kuchinskii_2017}
\begin{equation}
\chi_{\bm{q}_H}^\mathrm{(SC)} = \infty,
\label{eq:InFieldSusceptibilityDivergence}
\end{equation}
where ${q_{H}}^2 = 2 \pi \mu_0 H / \phi_0$~\footnote{Here we use  the following replacement: $q^2 \rightarrow {q_{H}}^2$, which should be valid if the magnetic field is low enough ($q_H$ is sufficiently smaller than both the inverse of the lattice constant and $q_{H=H_\mathrm{c2}(0)}$)~\cite{Dorin_1993, *Yanase_2000, Kuchinskii_2017}.
The reader can refer to Refs.~\cite{Lee_Payne_1971, *Lee_Payne_1972} for a more complete derivation of this replacement including the case of $H \sim H_\mathrm{c2} (0)$ within a continuum model.} and $\chi_{\bm{q}}^\mathrm{(SC)}$ is given in Eq.~\eqref{eq:SCSusceptibility}. Here $\mu_0$ is the vacuum permeability, and $\phi_0 = \pi \hbar / e$ is the flux quantum. $\chi_{\bm{q}_H}^\mathrm{(SC)}$ approximately describes the susceptibility for states with the lowest-Landau-level index and uniform in the $z$ direction.
Note that, as we show in Appendix~\ref{app:LLL}, Eq.~\eqref{eq:InFieldSusceptibilityDivergence} represents the transition point in the Gaussian approximation within the functional-integral formalism.
As for a free-particle number equation to determine the chemical potential, we adopt Eq.~\eqref{eq:NumberEqFree} since we neglect the Landau quantization of the electron kinetic energy. Therefore, we combine Eq.~\eqref{eq:InFieldSusceptibilityDivergence} with Eq.~\eqref{eq:NumberEqFree} to estimate $H^*$. The curve $(T, H^* (T))$ merges into $(T^*, 0)$ in the low-field limit; thus $H^*$ can be regarded as a natural extension of $T^*$ to the finite-field region.

The vortex-liquid formation field $H_\mathrm{c2}$ is estimated in a similar way to the calculation of $T_\mathrm{c}$. Since we focus on systems with $U < U_0$ (see the green and yellow dotted lines in Fig.~\ref{fig:muc-U}), where the decrease in $\mu$ is not so large and the $T$ dependence of $\mu$ is not so important, we simply approximate
\begin{equation}
\mu (T, H) \simeq \mu(T_\mathrm{c}, 0),
\label{eq:ChemicalPotentialInField}
\end{equation}
where $\mu (T_\mathrm{c}, 0)$ is obtained within the T-matrix approximation (see Sec.~\ref{sec:ZeroField} and Fig.~\ref{fig:muc-U}).
This approximation is correct at least in the weak-coupling BCS side since $\mu$ is almost fixed regardless of temperature~\footnote{See, e.g., Fig.~2 in Ref.~\cite{Kuchinskii_2017} for the temperature dependence of the chemical potential determined in the fluctuation region near $T_\mathrm{c}$. Note that, to examine the fluctuation-induced transition and crossover lines, we only have to construct the GL functional in the fluctuation region.}, and we believe that this approximation is a first step to consider magnetic-field effects in the case with strong attractive interaction.
After we replace $\mu (T, H)$ with $\mu(T_\mathrm{c}, 0)$, we solve Eq.~\eqref{eq:InFieldSusceptibilityDivergence} to estimate $H_\mathrm{c2}$. Similar to the case of $H^*$, the curve ($T, H_\mathrm{c2} (T)$) merges into ($T_\mathrm{c}, 0$) in the low-field limit; thus $H_\mathrm{c2}$ can be understood as an extension of $T_\mathrm{c}$ to the finite-field region.

Regarding the vortex-lattice-formation field $H_\mathrm{melt}$, we apply an analysis based on the GL functional~\cite{Ikeda_1990, *Ikeda_1992} in the lowest-Landau-level approximation~\cite{Eilenberger_1967, Moore_1989}, which is valid closer to the $H_\mathrm{c2}$ line~\cite{Moore_1989, Tesanovic_Xing_1991, *Tesanovic_1991}. First, as explained in Appendix~\ref{app:GL}, we derive the zero-field GL functional $\mathcal{F}_\mathrm{GL}$:
\begin{equation}
\mathcal{F}_\mathrm{GL} = \sum_{\bm{q}} T \left( 1 - U \chi_{\bm{q}}^{(0)} (0) \right) |a_{\bm{q}}|^2 + \frac{\beta}{2} \sum_i |a_i|^4.
\label{eq:GLFunctional}
\end{equation}
Here, $a_i$ is the order-parameter field defined on the lattice sites, $a_{\bm{q}}$ is its Fourier transformation satisfying $a_i = M^{-1/2} \sum_{\bm{q}} \exp (\mathrm{i} \bm{q} \cdot \bm{r}_i) a_{\bm{q}}$, and the coefficient $\beta$ is given as
\begin{equation}
\beta = \frac{T^3 U^2}{M} \sum_{\bm{k}, n} \left| G_{\bm{k}}^{(0)} (\mathrm{i} \varepsilon_n) \right|^4.
\label{Eq:beta}
\end{equation}
As shown in Appendix~\ref{app:LLL}, by applying the lowest-Landau-level approximation to Eq.~\eqref{eq:GLFunctional} with replacement of the momentum in the $x$-$y$ plane by $\bm{q}_H$ consistently with Eq.~\eqref{eq:InFieldSusceptibilityDivergence} and using the gradient expansion in the $z$ direction, we obtain
\begin{equation}
\mathcal{F}_\mathrm{GL} \simeq \int \mathrm{d}^3 \bm{r} \left[ \left( \alpha_{\bm{q}_H} |\psi (\bm{r})|^2 + \gamma |\partial_z \psi (\bm{r})|^2 \right) + \frac{\beta}{2} |\psi (\bm{r})|^4 \right],
\label{eq:GLFunctionalContinuous}
\end{equation}
where the order-parameter field $\psi (\bm{r})$ involves only the lowest-Landau-level modes in the $x$-$y$ plane. The coefficients are given as follows:
\begin{equation}
\alpha_{\bm{q}_H} = T \left( 1 - U \chi_{\bm{q}_H}^{(0)} (0) \right),
\label{Eq:AlphaqH}
\end{equation}
and
\begin{align}
\gamma =& - \frac{T^2 U t}{M} \sum_{\bm{k}, n} \left[ G_{\bm{k}}^{(0)} (\mathrm{i} \varepsilon_n) \right]^2 G_{\bm{-k}}^{(0)} (- \mathrm{i} \varepsilon_n) \nonumber \\
&\times \left[ \cos k_z + 4 t G_{\bm{k}}^{(0)} (\mathrm{i} \varepsilon_n) \sin^2 k_z \right]^2.
\label{Eq:gamma}
\end{align}
As shown in Appendix~\ref{app:Hmelt}, based on Eq.~\eqref{eq:GLFunctionalContinuous}, the vortex-lattice-formation field $H_\mathrm{melt}$ is approximately calculated by solving the following equation:
\begin{equation}
\frac{T}{4 \pi \sqrt{\rho_\mathrm{s} c_{66}}} = \frac{c^2}{h}.
\label{eq:VortexLatticeFormastionField}
\end{equation}
Here, $h = 2 \pi \mu_0 H / \phi_0$ is a dimensionless magnetic field (note that the lattice constant is set to unity), and $c = \mathcal{O} (10^{-1})$ is a phenomenological parameter~\cite{Bennemann_2008, Moore_1989}. Also, $c_{66}$ and $\rho_\mathrm{s}$ represent the shear modulus of the vortex lattice and the superfluid density defined \textit{along} the magnetic field, respectively (see Appendix~\ref{app:Hmelt}):
\begin{equation}
c_{66} = \frac{2 \gamma_\mathrm{A} |\alpha_{\bm{q}_H}|^2}{{\beta_\mathrm{A}}^2 \beta},
\end{equation}
and
\begin{equation}
\rho_\mathrm{s} = \frac{2 |\alpha_{\bm{q}_H}| \gamma}{\beta_\mathrm{A} \beta}
\end{equation}
with numerical factors related to the triangular vortex-lattice structure: $\beta_\mathrm{A} \simeq 1.16$ and $\gamma_\mathrm{A} \simeq 0.119$.
Note that $\beta$ and $\gamma$ appearing in these expressions are given in Eqs.~\eqref{Eq:beta} and \eqref{Eq:gamma}, respectively.
To obtain $H_\mathrm{melt}$, we solve Eq.~\eqref{eq:VortexLatticeFormastionField} in combination with the approximated chemical potential [Eq.~\eqref{eq:ChemicalPotentialInField}].

\section{Field--temperature phase diagram}
\label{sec:FieldTemperaturePhaseDiagram}

Based on numerically calculated $H^*$, $H_\mathrm{c2}$, and $H_\mathrm{melt}$, we obtain typical $H$-$T$ phase diagrams (Fig.~\ref{fig:H-T}). Since our purpose is to investigate qualitative features of the $H$-$T$ phase diagram, we fix the phenomenological parameter to estimate $H_\mathrm{melt}$ as $c = 0.5$ throughout our calculation. A slight change in $c$ does not affect the qualitative features.
The value of $c$ may be phenomenologically determined by comparing the resulting phase diagram with certain experiments or could be derived from a more complete theory.
Figures~\ref{fig:H-T}(a) and (b) show the weak-interaction ($U / t = 2.57$) and strong-interaction ($U / t = 5.14$) cases with lower density ($n = 0.2$), respectively. Comparing Figs.~\ref{fig:H-T}(a) and (b), we can see that the vortex-liquid region between $H_\mathrm{c2}$ and $H_\mathrm{melt}$, as well as the preformed-pair region between $H_\mathrm{c2}$ and $H^*$, becomes broader as the interaction becomes stronger; therefore, a strong attractive interaction stabilizes both the vortex-liquid and the preformed-pair regions.

Let us consider physical reasons why both the vortex-liquid and preformed-pair states are stabilized by a strong attractive interaction. First, the stabilization of the preformed-pair state can be understood in the same way as the zero-field case: a strong attractive interaction makes it easy to create non-condensed pairs, or preformed pairs~\cite{Randeria_Taylor_2014}. Second, the stabilization of the vortex-liquid region can be understood based on the superconducting-fluctuation strength: as the attractive interaction gets stronger toward the BCS-BEC crossover regime, the fluctuation becomes more significant~\cite{Debelhoir_Dupuis_2016, Adachi_Ikeda_2017}, and thus the vortex-liquid region becomes wider.

Figures~\ref{fig:H-T}(c) and (d) show the obtained phase diagrams with higher density ($n = 0.5$). Similar to the case with lower density ($n = 0.2$), we can see that both the vortex-liquid and the preformed-pair regions are stabilized when the interaction is strong. Moreover, comparing the higher density case [Figs.~\ref{fig:H-T}(a) and (b)] with the lower density case [Figs.~\ref{fig:H-T}(c) and (d)], we can see that the vortex-liquid region is broader while the preformed-pair region is narrower when the density is higher. From this result, we conclude that the particle density, in addition to interaction strength, is an important factor in determining the resultant $H$-$T$ phase diagram.

Here, we point out that keeping only the LLL modes among various order parameter's spatial variations is an approach from the weak fluctuation in the following sense: it is clear that, in the weak-field limit, the LLL mode vanishes so that the fluctuation-induced downward shift of $T_\mathrm{c}$ in zero field, $\Delta T_\mathrm{c}(0)$, cannot be described within the present approach. To describe $\Delta T_\mathrm{c}(0)$, it is necessary to incorporate the higher-Landau-level (HLL) modes in our calculation. In fact, the HLL modes incorporating the vortex-loop fluctuations~\cite{Ikeda_1995, Tesanovic_1999} should lead to not only $\Delta T_\mathrm{c}(0)$ and a shift of the $H_\mathrm{c2}(T)$ line in low fields accompanying it but also a downward shift of $H_\mathrm{melt}(T)$ and a change of its temperature dependence in low enough fields. Although such effects have been omitted in the present LLL approach, this simplification is not essential to our purpose here of understanding a qualitative picture of the $H$-$T$ phase diagram in superconductors with a strong pairing interaction.

So far, we consider the $H$-$T$ phase diagram based on the T-matrix approximation.
Beyond the T-matrix approximation used here, several kinds of more sophisticated approximations such as the self-consistent T-matrix approximation~\cite{Chen_Stajic_2005, Yanase_Yamada_1999, Haussmann_1994} have been applied to discuss the $U$-$T$ phase diagram, thermodynamic quantities, and others.
Here we briefly discuss what can occur if we use the self-consistent T-matrix approximation, instead of the method used in this study, to examine the $H$-$T$ phase diagram.
With the use of the self-consistent calculation, it is known that the pseudogap created by the preformed pair can be reflected in the electronic state through the self-energy effect, which reduces further the transition temperature $T_\mathrm{c}$~\cite{Yanase_Yamada_1999, Haussmann_1994}.
In the same way, $H_\mathrm{c2} (T)$ and $H_\mathrm{melt} (T)$ are expected to be further lowered; therefore, the separation between $H^* (T)$ and $H_\mathrm{c2} (T)$ as well as that between $H^* (T)$ and $H_\mathrm{melt} (T)$ will become more prominent as $U$ gets larger, compared with the results shown in Fig.~\ref{fig:H-T}.
Nonetheless, we believe that the qualitative features obtained in this study can be seen even if we apply the above-mentioned self-consistent calculation.

\section{Conclusion}

To obtain typical $H$-$T$ phase diagrams in electron systems with strong attractive interaction, we estimate the pair-formation field $H^*$, the vortex-liquid-formation field $H_\mathrm{c2}$, and the vortex-lattice-formation field $H_\mathrm{melt}$. Based on numerical calculations, we find that a strong attractive interaction can stabilize both the vortex-liquid and the preformed-pair regions. In addition, we point out that the particle density also influences the resultant phase diagram.

In the preformed-pair and vortex-liquid regions stabilized by strong attractive interaction, thermodynamic and transport properties are expected to be characteristic. In particular, the Hall conductivity in the vortex-liquid region can be enhanced by superconducting-fluctuation effecs~\cite{Fukuyama_1971, Aronov_1995} since the dynamics of the superconducting order parameter can involve a larger propagating part when the interaction is stronger~\cite{Melo_Randeria_1993}.

In the end of this paper, we discuss the $H$-$T$ phase diagram in FeSe suggested by several experiments~\cite{Kasahara_Watashige_2014, Kasahara_Yamashita_2016, Shi_Arai_2017}.
We should remark that FeSe has a two-band structure consisting of an electron Fermi surface and a hole one, so that our analysis on the single-band Hubbard model will not completely describe the physical properties of FeSe.
Nevertheless, let us compare our numerical results with the experimental data observed in FeSe.
Also, we do not comment on the high-field low-temperature phase (``$B$-phase'') proposed in Ref.~\cite{Kasahara_Watashige_2014} since in our calculation we do not take into account the Zeeman coupling of magnetic field, which may be important in the high-field low-temperature region.

First, a large pseudogap region above $H_\mathrm{c2}$ in the $H$-$T$ plane is suggested in Ref.~\cite{Shi_Arai_2017}. If we assume that the pseudogap is caused by the preformed pair~\cite{Chen_Stajic_2005, Tsuchiya_Watanabe_2009}, we can interpret the observed pseudogap region as the preformed-pair region stabilized by a strong attractive interaction as in Fig.~\ref{fig:H-T}(b). Second, a crossing of magnetization curves~\cite{Li_Suenaga_1992, *Tesanovic_Xing_1992} is observed in Ref.~\cite{Kasahara_Yamashita_2016}. This crossing can be understood as caused by a strong attractive interaction~\citep{Adachi_Ikeda_2017} in the vortex-liquid region stabilized also by the strong attractive interaction. Third, the Hall, Seebeck, and Nernst coefficients have shown their maximum or minimum near a temperature $T \sim 2 T_\mathrm{c}$ with weak dependence on $H$~\cite{Kasahara_Yamashita_2016}. Though a strong attractive interaction may be related to this behavior, the detailed electronic structure~\cite{Kasahara_Watashige_2014, Terashima_Kikugawa_2014, Sprau_Kostin_2017} should be taken into account to discuss such transport phenomena since FeSe is an almost compensated semimetal~\cite{Kasahara_Watashige_2014} and compensation of electron and hole carriers can make the sign of transport coefficients, such as the Hall coefficient, subtle.

In addition, we discuss the resistive vanishing in FeSe in finite fields. As stressed in the present work as well as Ref.~\cite{Adachi_Ikeda_2018}, a broad preformed-pair region is expected, as in Fig.~\ref{fig:H-T}(b), to lie above the nominal $H_\mathrm{c2}(T)$ curve in FeSe. If so, the fact~\cite{Kasahara_Watashige_2014, Kasahara_Yamashita_2016} that the vortex-liquid region is relatively narrow in the experimental phase diagram of FeSe needs to be clarified. This discrepancy may be due to the fact that the resistivity vanishes at a much higher temperature than $H_\mathrm{melt}(T)$ defined in clean limit. This possibility occurs when the resistivity vanishes through a vortex-glass transition due to the vortex pinnings to columnar defects or correlated defects~\cite{Nelson_1993, Ikeda_2000}. Another possibility is that the vortex-liquid region has estimated to be much narrower from the resistivity data than the actual one. This may occur when the quantum fluctuation neglected in the present study is not negligible~\cite{Ikeda_2003}. If this scenario is true, the resistivity is insensitive to the position of the actual $H_\mathrm{c2}$ and, upon cooling, begins to vanish close to a vortex-glass transition, which lies near $H_\mathrm{melt}$ and much below the actual $H_\mathrm{c2}$.

As another possible scenario to explain why the vortex-liquid region is estimated to be relatively narrow in FeSe, let us consider the two-band structure characteristic of FeSe~\cite{Kasahara_Watashige_2014, Terashima_Kikugawa_2014, Sprau_Kostin_2017}. If a strong attractive interaction is present in one of these bands while a weak attractive interaction exists in another band, the vortices due to the former band can be pinned by the vortex lattice generated by the latter band. If this is true, the vortex-liquid region can become relatively narrow compared to the case considered in the present work where only a single band with strong attractive interaction exists. This possibility will be examined in details elsewhere.

\textit{Acknowledgments}. The present research was supported by JSPS KAKENHI [Grants No.~16K05444 and No.~17J03883]. K.~A. also thanks JSPS for support from a Research Fellowship for Young Scientists.

\appendix

\begin{figure}[t]
\includegraphics[scale=0.65]{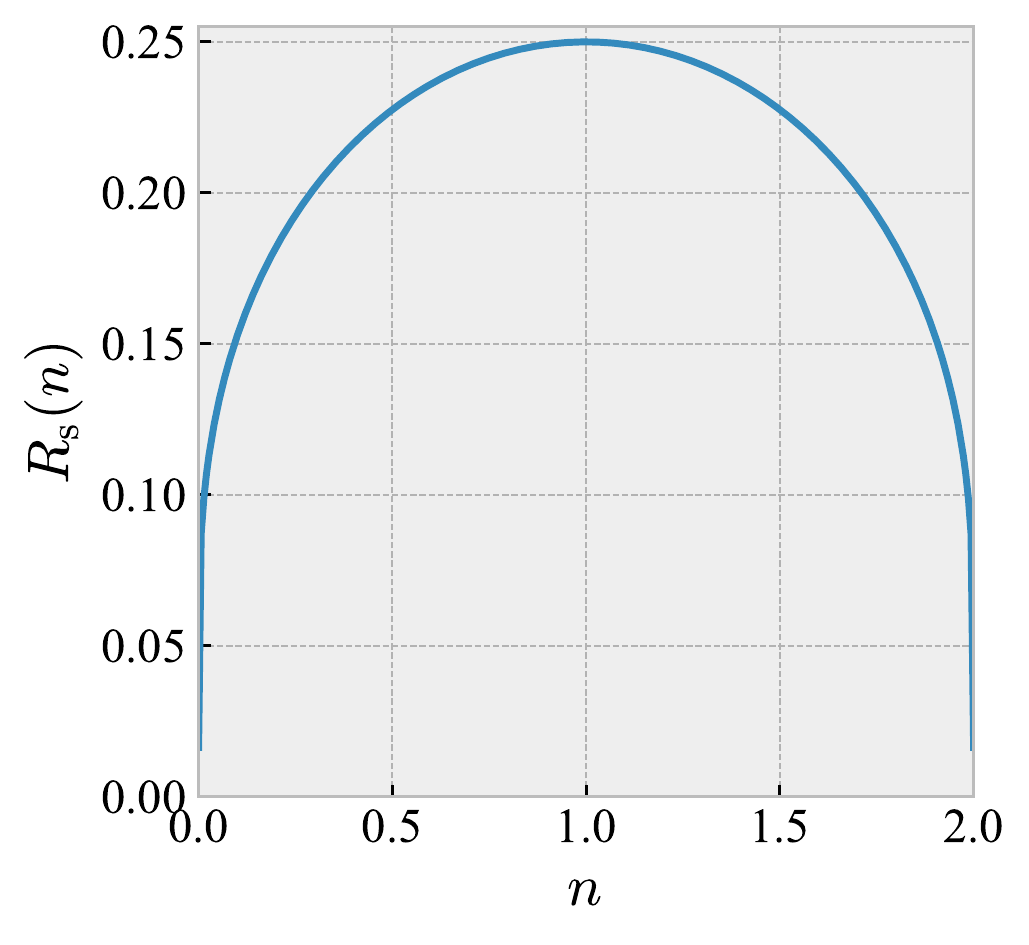}
\caption{Functional form of $R_\mathrm{s}(n)$, the ratio of $T^*$ to $E_\mathrm{gap}$ in the strong-coupling limit [Eq.~\eqref{Eq:RatioStrongCouplingLimit}].}
\label{Fig:Func_n}
\end{figure}

\section{Comparison between pair-formation temperature and zero-temperature excitation energy gap}
\label{app:Delta}

In this section, let us compare the pair-formation temperature $T^*$ with the zero-temperature excitation energy gap $E_\mathrm{gap}$ to show that $T^*$ and $E_\mathrm{gap}$ are of the same order of magnitude ($0.1 \lesssim T^* / E_\mathrm{gap} \lesssim 0.5$) regardless of the interaction strength $U$ across a broad range of the particle density $n$ ($0.015 \lesssim n \lesssim 1.985$).
Based on this fact, as discussed in the following, $T^*$ can be interpreted as the pair-formation temperature.

The pair-formation temperature $T^*$ is calculated from Eqs.~\eqref{eq:UniformSusceptibilityDivergence} and \eqref{eq:NumberEqFree}, or rewritten as
\begin{equation}
\frac{1}{M} \sum_{\bm{k}} \frac{\tanh \left[ \left( \epsilon_{\bm{k}} - \mu \right) / (2 T) \right]}{2 \left( \epsilon_{\bm{k}} - \mu \right)} = \frac{1}{U},
\label{Eq:GapEqTstar}
\end{equation}
and
\begin{equation}
\frac{1}{M} \sum_{\bm{k}} \left[ 1 - \tanh \left( \frac{\epsilon_{\bm{k}} - \mu}{2 T} \right) \right] = n.
\label{Eq:NumberEqTstar}
\end{equation}

To estimate the zero-temperature excitation energy, we need to calculate the zero-temperature superconducting gap amplitude $\Delta_0$.
$\Delta_0$ can be estimated through the BCS-BEC crossover with the use of the mean-field approximation or variational BCS wave function~\cite{Nozieres_Schmitt-Rink_1985, Micnas_1990, Chen_Stajic_2005}.
The resulting equations to calculate $\Delta_0$ is given as the gap equation
\begin{equation}
\frac{1}{M} \sum_{\bm{k}} \frac{1}{2 \sqrt{(\epsilon_{\bm{k}} - \widetilde{\mu})^2 + {\Delta_0}^2}} = \frac{1}{U},
\label{Eq:GapEqDelta}
\end{equation}
and the particle-number equation
\begin{equation}
\frac{1}{M} \sum_{\bm{k}} \left( 1 - \frac{\epsilon_{\bm{k}} - \widetilde{\mu}}{\sqrt{(\epsilon_{\bm{k}} - \widetilde{\mu})^2 + {\Delta_0}^2}} \right) = n,
\label{Eq:NumberEqDelta}
\end{equation}
where the Hartree shift is incorporated by redefining the chemical potential as $\widetilde{\mu} = \mu + n U / 2$~\footnote{Interestingly, in the equations for $T^*$ [Eqs.~(A1) and (A2)], replacing $\tanh [(\epsilon_{\bm{k}} - \mu) / (2 T)]$ with $(\epsilon_{\bm{k}} - \tilde{\mu}) / \sqrt{(\epsilon_{\bm{k}} - \tilde{\mu})^2 + {\Delta_0}^2}$ as well as $\epsilon_{\bm{k}} - \mu$ with $\epsilon_{\bm{k}} - \tilde{\mu}$, we exactly obtain the equations for $\Delta_0$ [Eqs.~(A3) and (A4)].
Moreover, as pointed out in, e.g., Ref.~\cite{Tinkham_1996}, the functional form of $\tanh (x / a)$ and that of $x / \sqrt{x^2 + a^2}$ ($a > 0$) are quite similar to each other; thus, for given values of $U$ and $n$, the obtained $T^*$ and $\Delta_0$ are expected to be of the same order of magnitude [$T^* / \Delta_0 = \mathcal{O} (1)$]. We can numerically show this fact across a broad range of $U$ and $n$.}.

\begin{figure*}[t]
\includegraphics[scale=0.7]{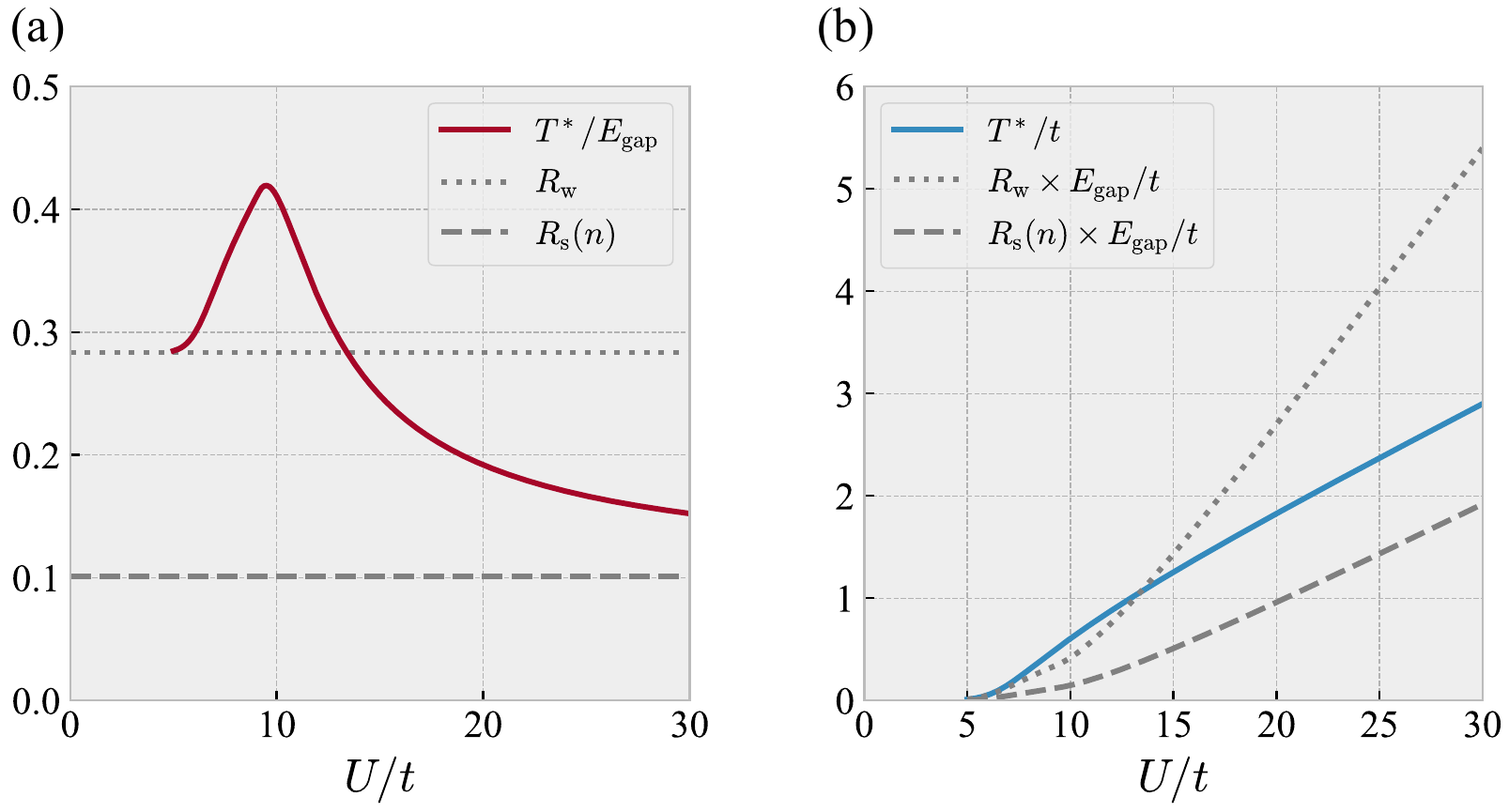}
\caption{(a) The ratio of the pair-formation temperature $T^*$ to the zero-temperature excitation energy gap $E_\mathrm{gap}$ as a function of the interaction strength $U$ for the particle density $n = 0.015$ (red solid line).
The asymptotic ratio in the weak-coupling limit $R_\mathrm{w} \simeq 0.283$ (grey dotted line) and that in the strong-coupling limit $R_\mathrm{s} (n = 0.015) \simeq 0.101$ (grey dashed line) are also shown [see Eqs.~\eqref{Eq:RatioWeakCouplingLimit} and \eqref{Eq:RatioStrongCouplingLimit} for the definitions of $R_\mathrm{w}$ and $R_\mathrm{s} (n)$, respectively].
(b) $U$ dependence of $T^*$ (blue solid line).
The asymptotic $U$ dependence of $T^*$ in the weak-coupling limit $R_\mathrm{w} \times E_{\mathrm{gap}}$ (grey dotted line) and that in the strong-coupling limit $R_\mathrm{s} (n = 0.015) \times E_\mathrm{gap}$ (grey dashed line) are also shown.}
\label{Fig:RatioTstarEgap_n0015}
\end{figure*}

\begin{figure*}[t]
\includegraphics[scale=0.7]{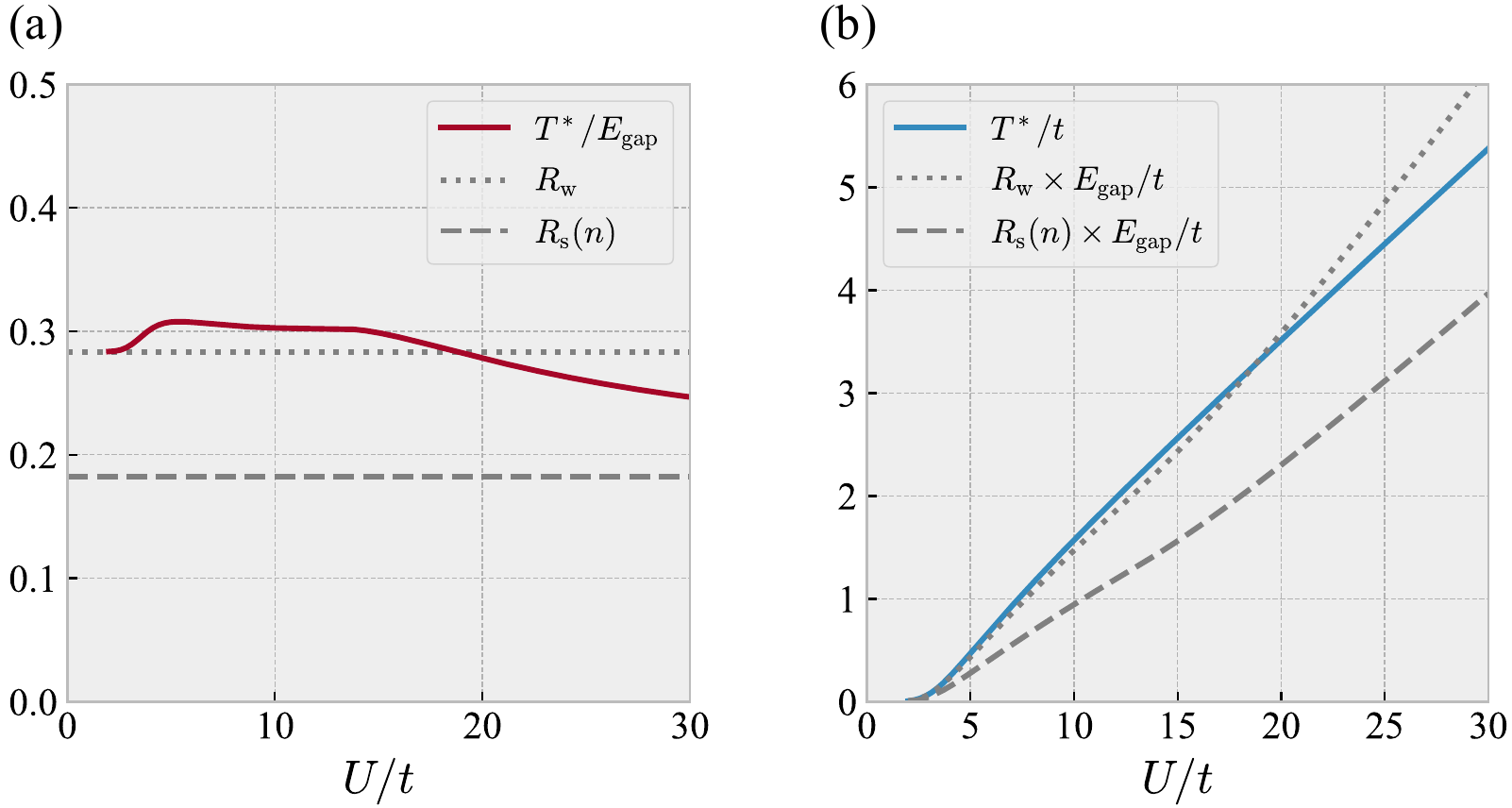}
\caption{$U$ dependence of (a) $T^* / E_\mathrm{gap}$ and (b) $T^*$ corresponding to Fig.~\ref{Fig:RatioTstarEgap_n0015} for the case of $n = 0.2$.
The meaning of each curve is the same as used in Fig.~\ref{Fig:RatioTstarEgap_n0015}.
Note that $R_\mathrm{w} \simeq 0.283$ and $R_\mathrm{s} (n = 0.2) \simeq 0.182$.}
\label{Fig:RatioTstarEgap_n02}
\end{figure*}

The one-particle excitation energy $E_{\bm{k}}$~\cite{Tinkham_1996} is given as~\cite{Micnas_1990}
\begin{equation}
E_{\bm{k}} = \sqrt{(\epsilon_{\bm{k}} - \widetilde{\mu})^2 + {\Delta_0}^2},
\end{equation}
and the minimum excitation energy $E_\mathrm{gap}$, i.e., the lowest energy to break an electron pair without changing the total number of particles, is given as~\cite{Micnas_1990}
\begin{equation}
E_\mathrm{gap} = 2 \min_{\bm{k}} E_{\bm{k}}.
\end{equation}
Therefore, for the weak-coupling side satisfying $|\widetilde{\mu}| < 6t$ ($=$ half of the band width), $E_\mathrm{gap}$ is given by $2 \Delta_0$; on the other hand, for the strong-coupling side satisfying $|\widetilde{\mu}| > 6t$, $E_\mathrm{gap}$ is given by $2 \sqrt{(|\widetilde{\mu}| - 6t)^2 + {\Delta_0}^2}$.

To investigate the ratio of $T^*$ to $E_\mathrm{gap}$, let us first consider the weak-coupling limit ($U / t \ll 1$), where $E_\mathrm{gap} = 2 \Delta_0$.
In this limit, $T^*$ is essentially equivalent to the pair-condensation temperature $T_\mathrm{c}$, and the well-known BCS relation for the $s$-wave superconductivity~\cite{Tinkham_1996} is applicable:
\begin{equation}
\frac{T^*}{E_\mathrm{gap}} \simeq \frac{T_\mathrm{c}}{E_\mathrm{gap}} \simeq \frac{\mathrm{e}^{\gamma_\mathrm{E}}}{2 \pi} \equiv R_\mathrm{w} \simeq 0.283 \ \ (\text{for } U / t \ll 1),
\label{Eq:RatioWeakCouplingLimit}
\end{equation}
where $\gamma_\mathrm{E} \simeq 0.577$ is the Euler's constant, and $R_\mathrm{w}$ represents $T^* / E_\mathrm{gap}$ in the weak-coupling limit.
Next, let us consider the strong-coupling limit ($U / t \gg 1$), where $E_\mathrm{gap} = 2 \sqrt{(|\widetilde{\mu}| - 6t)^2 + {\Delta_0}^2}$.
In this limit, we can obtain from Eqs.~\eqref{Eq:GapEqTstar} and \eqref{Eq:NumberEqTstar}
\begin{equation}
T^* \simeq \frac{1 - n}{2 \ln (2 / n - 1)} U \ \ (\text{for } U / t \gg 1).
\end{equation}
On the other hand, in the same strong-coupling limit, we obtain from Eqs.~\eqref{Eq:GapEqDelta} and \eqref{Eq:NumberEqDelta}~\cite{Micnas_1990}
\begin{equation}
\Delta_0 \simeq \frac{\sqrt{n (2 - n)}}{2} U \ \ (\text{for } U / t \gg 1),
\end{equation}
as well as
\begin{equation}
\widetilde{\mu} \simeq \frac{n - 1}{2} U \ \ (\text{for } U / t \gg 1).
\end{equation}
Thus, we obtain the expression $E_\mathrm{gap} \simeq U$, where $U$ is just the binding energy of two particles in the strong-coupling limit~\cite{Micnas_1990}.
Therefore, in the strong-coupling limit, we see
\begin{equation}
\frac{T^*}{E_\mathrm{gap}} \simeq \frac{1 - n}{2 \ln (2 / n - 1)} \equiv R_\mathrm{s}(n) \ \ (\text{for } U / t \gg 1).
\label{Eq:RatioStrongCouplingLimit}
\end{equation}
Figure~\ref{Fig:Func_n} shows the $n$ dependence of $R_\mathrm{s}(n)$, which represents $T^* / E_\mathrm{gap}$ in the strong-coupling limit.
From Fig.~\ref{Fig:Func_n}, we may see that $0.1 \lesssim R_\mathrm{s} (n) \lesssim 0.25$ across a broad range of the particle density $n$ ($0.015 \lesssim n \lesssim 1.985$).

Based on the above consideration of the weak- and strong-coupling limits, we can expect that $T^* \sim E_\mathrm{gap}$ through the BCS-BEC crossover, or irrespective of the value of $U$, as long as the density $n$ is neither very low nor very high ($0.015 \lesssim n \lesssim 1.985$).
As examples, we present the numerically calculated $U$ dependence of $T^* / E_\mathrm{gap}$ for $n = 0.015$ in Fig.~\ref{Fig:RatioTstarEgap_n0015}(a) and $n = 0.2$ in Fig.~\ref{Fig:RatioTstarEgap_n02}(a) [for reference, the $U$ dependence of $T^*$ is also shown in Fig.~\ref{Fig:RatioTstarEgap_n0015}(b) for $n = 0.015$ and Fig.~\ref{Fig:RatioTstarEgap_n02}(b) for $n = 0.2$].
These figures show that $0.1 \lesssim T^* / E_\mathrm{gap} \lesssim 0.5$ is satisfied in the BCS-BEC crossover regime as well as in the weak- and strong-coupling regimes.

To sum up, we obtain $0.1 \lesssim T^* / E_\mathrm{gap} \lesssim 0.5$ regardless of the interaction strength $U$ across a broad range of the particle density $n$ ($0.015 \lesssim n \lesssim 1.985$).
Since the zero-temperature excitation energy gap $E_\mathrm{gap}$ can be interpreted as a typical energy scale to break an electron pair, $T^*$ also can be regarded as a typical temperature scale to break a pair; therefore, $T^*$ is expected to represent the pair-breaking, or pair-formation, temperature.

\section{Derivation of Ginzuburg-Landau functional}
\label{app:GL}

Here we derive the zero-field GL functional given by Eq.~\eqref{eq:GLFunctional}. By using the functional integral representation~\cite{Melo_Randeria_1993, Liu_Zhai_2014, Debelhoir_Dupuis_2016}, we can formally rewrite the grand-canonical partition function $Z$ as
\begin{equation}
Z = \int \left[ \prod_{\bm{k}, \sigma, n} \mathrm{d} c^*_{\bm{k} \sigma} (\varepsilon_n) \mathrm{d} c_{\bm{k} \sigma} (\varepsilon_n) \right] \mathrm{e}^{-(S_0 + S_\mathrm{int})},
\end{equation}
where
\begin{equation}
S_0 = \frac{1}{T} \sum_{\bm{k}, \sigma, n} \left[ -G_{\bm{k}}^{(0)} (\mathrm{i} \varepsilon_n)^{-1} \right] c^*_{\bm{k} \sigma} (\varepsilon_n) c_{\bm{k} \sigma} (\varepsilon_n),
\end{equation}
\begin{equation}
S_\mathrm{int} = - \frac{U}{T M} \sum_{\bm{q}, m} \phi^*_{\bm{q}} (\omega_m) \phi_{\bm{q}} (\omega_m),
\end{equation}
and
\begin{equation}
\phi_{\bm{q}} (\omega_m) = \sum_{\bm{k}, n} c_{-\bm{k} \downarrow} (-\varepsilon_n) c_{\bm{k} + \bm{q} \uparrow} (\varepsilon_n + \omega_m).
\end{equation}
Here, $c_{\bm{k} \sigma} (\varepsilon_n)$ and $c^{*}_{\bm{k} \sigma} (\varepsilon_n)$ are the Grassmann numbers, and $G_{\bm{k}}^{(0)} (\mathrm{i} \varepsilon_n) = (\mathrm{i} \varepsilon_n - \epsilon_{\bm{k}} + \mu)^{-1}$ is the non-interacting Green's function.

Introducing the order-parameter field $a_{\bm{q}} (\omega_m)$ and $a^*_{\bm{q}} (\omega_m)$ with the Hubbard-Stratonovich transformation, we can obtain the following expression:
\begin{align}
\mathrm{e}^{-S_\mathrm{int}} =& \int \left[ \prod_{\bm{q}, m} \frac{\mathrm{d} a^*_{\bm{q}} (\omega_m) \mathrm{d} a_{\bm{q}} (\omega_m)}{\pi} \right] \mathrm{e}^{-\sum_{\bm{q}, m} |a_{\bm{q}} (\omega_m)|^2} \nonumber \\
& \times \mathrm{e}^{\sqrt{U / (T M)} \sum_{\bm{q}, m} [a^*_{\bm{q}} (\omega_m) \phi_{\bm{q}} (\omega_m) + \mathrm{c.c.}]}.
\end{align}
Using this expression, we can transform the partition function as
\begin{align}
\frac{Z}{Z_0} =& \langle \mathrm{e}^{-S_\mathrm{int}} \rangle_0 \nonumber \\
=& \int \left[ \prod_{\bm{q}, m} \frac{\mathrm{d} a^*_{\bm{q}} (\omega_m) \mathrm{d} a_{\bm{q}} (\omega_m)}{\pi} \right] \mathrm{e}^{-\sum_{\bm{q}, m} |a_{\bm{q}} (\omega_m)|^2} \nonumber \\
& \times \left< \mathrm{e}^{\sqrt{U / (T M)} \sum_{\bm{q}, m} [a^*_{\bm{q}} (\omega_m) \phi_{\bm{q}} (\omega_m) + \mathrm{c.c.}]} \right>_0.
\label{eq:PartitionFunction}
\end{align}
Here, $Z_0 = \int [ \prod_{\bm{k}, \sigma, n} \mathrm{d} c^*_{\bm{k} \sigma} (\varepsilon_n) \mathrm{d} c_{\bm{k} \sigma} (\varepsilon_n) ] \mathrm{e}^{-S_0}$ is the non-interacting partition function, and $\langle \cdots \rangle_0$ represents the grand-canonical ensemble average with respect to the non-interacting part $S_0$. Expanding the last term in Eq.~\eqref{eq:PartitionFunction} with respect to the order-parameter field $a_{\bm{q}} (\omega_m)$ and $a^*_{\bm{q}} (\omega_m)$ up to the fourth order and neglecting Bosonic quantum fluctuation, we can finally obtain the following form:
\begin{equation}
\frac{Z}{Z_0} \sim \int \left[ \prod_{\bm{q}} \frac{\mathrm{d} a^*_{\bm{q}} \mathrm{d} a_{\bm{q}}}{\pi} \right] \mathrm{e^{-\mathcal{F}_\mathrm{GL} / T}},
\end{equation}
where we write $a_{\bm{q}} = a_{\bm{q}} (0)$ for simplicity. Here, $\mathcal{F}_\mathrm{GL}$ is the GL functional, the explicit form of which is given as
\begin{equation}
\mathcal{F}_\mathrm{GL} = \sum_{\bm{q}} T \left[ 1 - U \chi_{\bm{q}}^{(0)} (0) \right] |a_{\bm{q}}|^2 + \frac{\beta}{2} \sum_i |a_i|^4,
\label{eq:GLFunctionalApp}
\end{equation}
where $a_i = M^{-1 / 2} \sum_{\bm{q}} \exp (\mathrm{i} \bm{q} \cdot \bm{r}_i) a_{\bm{q}}$ is the real-space order-parameter field,
\begin{equation}
\chi_{\bm{q}}^{(0)} (\mathrm{i} \omega_m) = \frac{T}{M} \sum_{\bm{k}, n} G_{\bm{k} + \bm{q}}^{(0)} (\mathrm{i} \varepsilon_n + \mathrm{i} \omega_m) G_{- \bm{k}}^{(0)} (- \mathrm{i} \varepsilon_n),
\end{equation}
and
\begin{equation}
\beta = \frac{T^3 U^2}{M} \sum_{\bm{k}, n} \left| G_{\bm{k}}^{(0)} (\mathrm{i} \varepsilon_n) \right|^4.
\end{equation}

\section{Lowest-Landau-level approximation of Ginzburg-Landau action}
\label{app:LLL}

In the following, we explain how we obtain the approximated expression of the GL functional [Eq.~\eqref{eq:GLFunctionalContinuous}]. Neglecting the Landau quantization of electrons, the external magnetic field affects the energy eigenstate of the order-parameter field $a_i$. At large length scales, the lattice structure is not important so that we can focus on the long-wavelength parts of $a_{i}$ and can replace $a_i$ defined on lattice with $\psi (\bm{r})$ defined in continuum space (note that the lattice constant is set to unity). Then, to perform our calculation in a finite magnetic field parallel to the $z$ axis, we can rewrite Eq.~\eqref{eq:GLFunctional} as
\begin{equation}
\mathcal{F}_\mathrm{GL} \simeq \int \mathrm{d}^3 \bm{r} \left( \psi^* \alpha_{\bm{Q}} \psi + \gamma |\partial_z \psi |^2 + \frac{\beta}{2} |\psi|^4 \right),
\label{eq:GLFunctionalLongWavelength}
\end{equation}
where
\begin{equation}
\alpha_{\bm{Q}} = T \left[ 1 - U \chi_{\bm{Q}}^{(0)} (0) \right]
\end{equation}
with $\bm{Q} = - \mathrm{i} \nabla_\perp + 2 \pi \bm{A} / \phi_0$ is the gauge-invariant gradient in the directions perpendicular to the field, and
\begin{align}
\gamma = - \frac{T^2 U t}{M} \sum_{\bm{k}, n} & \left[ G_{\bm{k}}^{(0)} (\mathrm{i} \varepsilon_n) \right]^2 G_{\bm{-k}}^{(0)} (- \mathrm{i} \varepsilon_n) \nonumber \\
&\times \left[ \cos k_z + 4 t G_{\bm{k}}^{(0)} (\mathrm{i} \varepsilon_n) \sin^2 k_z \right]^2.
\end{align}
Here we introduce magnetic-field effects through a minimal coupling of the vector potential $\bm{A} (\bm{r})$ to the order-parameter field $\psi (\bm{r})$.

To diagonalize the second-order terms of Eq.~\eqref{eq:GLFunctionalLongWavelength}, we expand the order-parameter field as
\begin{equation}
\psi (\bm{r}) = \sum_{N, n_\mathrm{d}, q_z} b_{N n_\mathrm{d} q_z} f_{N n_\mathrm{d}} (x, y) \frac{\mathrm{e}^{\mathrm{i} q_z z}}{\sqrt{L_z}},
\label{eq:OPExpansion}
\end{equation}
where $N$ is the Landau-level index, $n_\mathrm{d}$ is the degeneracy index for each Landau level with $(\mu_0 H L_x L_y / \phi_0)$-fold degeneracy, $q_z$ is the $z$-directional momentum, and $f_{N n_\mathrm{d}} (x, y)$ is the $N$th Landau-level eigenfunction [note that the lattice constant is unity so that $L_i = M_i$ ($i = x, y, z$)]. Though, in general, it is not clear whether the second-order terms of Eq.~\eqref{eq:GLFunctionalLongWavelength} are diagonalized with the bases appearing in Eq.~\eqref{eq:OPExpansion}, at least the lowest-order $Q^2$ terms are exactly diagonalized with these bases. Respecting this fact and substituting Eq.~\eqref{eq:OPExpansion} into Eq.~\eqref{eq:GLFunctionalLongWavelength}, we obtain the diagonalized second-order terms:
\begin{equation}
\mathcal{F}_\mathrm{GL} \simeq \sum_{N, n_\mathrm{d}, q_z} \left( \alpha_{\sqrt{2N + 1}\bm{q}_H} + \gamma {q_z}^2 \right) |b_{N n_\mathrm{d} q_z}|^2 + \int \mathrm{d}^3 \bm{r} \frac{\beta}{2} |\psi|^4,
\label{eq:GLFunctionalGradientExp}
\end{equation}
where ${q_H}^2 = l^{-1} = \sqrt{2 \pi \mu_0 H / \phi_0}$.
Therefore, through the Landau quantization of the order-parameter field, we basically replace squared gauge-invariant gradient $Q^2$ defined in the $x$-$y$ plane with discrete levels $(2 N + 1) / l^2$.

As far as we focus our attention on the region relatively near $H_\mathrm{c2} (T)$, we just take into account the contribution from the lowest Landau-level mode~\cite{Ikeda_1995, Moore_1989, Tesanovic_Xing_1991, *Tesanovic_1991}; then we can obtain from Eq.~\eqref{eq:GLFunctionalGradientExp} the following expression:
\begin{equation}
\mathcal{F}_\mathrm{GL} \simeq \sum_{n_\mathrm{d}, q_z} \left( \alpha_{\bm{q}_H} + \gamma {q_z}^2 \right) |b_{0 n_\mathrm{d} q_z}|^2 + \int \mathrm{d}^3 \bm{r} \frac{\beta}{2} |\psi|^4.
\label{Eq:FGL_var}
\end{equation}
From this representation, we can see that the transition point in the Gaussian approximation is given as the vanishing point of the second-order mass term, i.e., $\alpha_{\bm{q}_H} = 0$, which can be rewritten as $\chi_{\bm{q}_H}^\mathrm{(SC)} = \infty$ [Eq.~\eqref{eq:InFieldSusceptibilityDivergence}] based on Eqs.~\eqref{eq:SCSusceptibility} and \eqref{Eq:AlphaqH}.
Conversely applying the expansion of the order-parameter field [Eq.~\eqref{eq:OPExpansion}] to Eq.~\eqref{Eq:FGL_var} as well as considering only $N = 0$ mode, we finally obtain
\begin{equation}
\mathcal{F}_\mathrm{GL} \simeq \int \mathrm{d}^3 \bm{r} \left[ \left( \alpha_{\bm{q}_H} |\psi|^2 + \gamma |\partial_z \psi|^2 \right) + \frac{\beta}{2} |\psi|^4 \right],
\end{equation}
where $\psi (\bm{r})$ only involves the lowest Landau-level mode ($N = 0$).

\begin{figure}[tbp]
\includegraphics[scale=1]{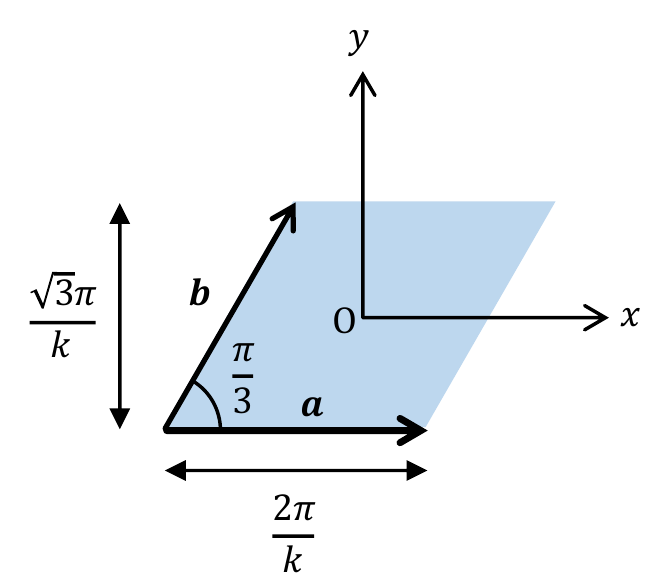}
\caption{Schematic figure of a unit cell of the triangular vortex lattice (blue area). Primitive lattice vectors ($\bm{a}$ and $\bm{b}$) as well as the size of the unit cell are shown. Note that one quantum flux penetrates one unit cell [$(\sqrt{3} \pi / k) \cdot (2 \pi / k) = 2 \pi l^2 = \phi_0 / (\mu_0 H)$].}
\label{fig:UnitCell}
\end{figure}

\section{Derivation of vortex-lattice-formation field}
\label{app:Hmelt}

In the following, we explain how we estimate the vortex-lattice-formation field $H_\mathrm{melt}$ and obtain Eq.~\eqref{eq:VortexLatticeFormastionField} starting with Eq.~\eqref{eq:GLFunctionalContinuous}. Since the mean-field solution minimizing Eq.~\eqref{eq:GLFunctionalContinuous} is given by the triangular vortex-lattice state, we consider the Gaussian fluctuation around the triangular vortex-lattice state~\cite{Ikeda_1990, *Ikeda_1992} within the lowest-Landau-level approximation~\cite{Eilenberger_1967, Moore_1989} and then apply the Lindemann criterion to estimate $H_\mathrm{melt}$~\cite{Moore_1989}, at which the first-order melting transition to the vortex-liquid state occurs. Since our formulation is basically based on Refs.~\cite{Eilenberger_1967, Moore_1989}, we here just present an overview. In the following, we assume the Landau gauge $\bm{A} (\bm{r}) = - \mu_0 H y \widehat{x}$. In this Appendix, $\bm{r}$ denotes a coordinate vector $x \widehat{x} + y \widehat{y}$ in the $x$-$y$ plane.

As a complete orthonormal set of bases diagonalizing the second-order terms of Eq.~\eqref{eq:GLFunctionalContinuous}, we consider a set of triangular vortex-lattice states with $z$-directional modulation:
\begin{equation}
\left\{ \varphi (\bm{r} | \bm{r}_0) \frac{\mathrm{e}^{\mathrm{i} q_z z}}{\sqrt{L_z}} \right\}_{\bm{r}_0, q_z},
\end{equation}
where $\{ \varphi (\bm{r} | \bm{r}_0) \}$ represents a two-dimensional triangular vortex lattice with a unit cell shown in Fig.~\ref{fig:UnitCell}, and the position of the vortices is related to $\bm{r}_0$:
\begin{equation}
\bm{r}_0 = x_0 \widehat{x} + y_0 \widehat{y} = \left( \frac{2 \pi l^2}{L_y} n_x + \frac{2 \pi l^2}{\sqrt{3} L_x} n_y \right) \widehat{x} + \frac{2 \pi l^2}{L_x} n_y \widehat{y}.
\end{equation}
Here $l = \sqrt{\phi_0 / (2 \pi \mu_0 H)}$ is the magnetic length. The degeneracy indices of the lowest Landau level, $n_x$ and $n_y$, satisfy
\begin{equation}
n_x \in \left[ -\frac{L_y}{2 k l^2}, \ \frac{L_y}{2 k l^2} \right), \ \ n_y \in \left[ - \frac{\sqrt{3} L_x}{4 k l^2}, \ \frac{\sqrt{3} L_x}{4 k l^2} \right)
\end{equation}
with $k = \sqrt{\sqrt{3} \pi} / l$. We note that the degeneracy of the lowest Landau level can be calculated as $[L_y / (k l^2)] \cdot [\sqrt{3} L_y / (2 k l^2)] = L_x L_y / (2 \pi l^2) = \mu_0 H L_x L_y / \phi_0$. The domain of $\bm{r}_0$ is equivalent to the unit cell shown in Fig.~\ref{fig:UnitCell}. As shown in the following, functions $\{ \varphi (\bm{r} | \bm{r}_0) \}_{\bm{r}_0}$ with $\bm{r}_0$ out of the unit cell are linearly dependent on those with $\bm{r}_0$ within the unit cell.

The specific form of the eigenfunctions $\{ \varphi (\bm{r} | \bm{r}_0) \}$ is given as
\begin{equation}
\varphi (\bm{r} | \bm{r}_0) = \mathrm{e}^{- \mathrm{i} y_0 x / l^2} \varphi (\bm{r} + \bm{r}_0 | \bm{0}),
\label{eq:varphi_r_r0}
\end{equation}
and
\begin{equation}
\varphi (\bm{r} |\bm{0}) = \frac{3^{1 / 8}}{\sqrt{L_x L_y}} \sum_{n = -\infty}^{\infty} \mathrm{e}^{\mathrm{i} k n x - \mathrm{i} \pi n^2 / 2 - (y - k l^2 n)^2 / (2 l^2)}.
\label{eq:varphi_r_0}
\end{equation}
Defining primitive lattice vectors $\bm{a} = (2 \pi / k) \widehat{x}$ and $\bm{b} = (\pi / k) \widehat{x} + (\sqrt{3} \pi / k) \widehat{y}$ as shown in Fig.~\ref{fig:UnitCell}, we obtain from Eq.~\eqref{eq:varphi_r_0} the following (quasi)periodicity of $\varphi (\bm{r} | \bm{0})$:
\begin{equation}
\left\{
\begin{aligned}
& \varphi (\bm{r} + \bm{a} | \bm{0}) = \varphi (\bm{r} | \bm{0}) \\
& \varphi (\bm{r} + \bm{b} | \bm{0}) = \mathrm{i} \mathrm{e}^{\mathrm{i} k x} \varphi (\bm{r} | \bm{0}),
\end{aligned}
\right.
\label{eq:TranslationPrimitive}
\end{equation}
As for a general lattice vector $\bm{R} = m_a \bm{a} + m_b \bm{b}$, we can show from Eq.~\eqref{eq:TranslationPrimitive} the following quasiperiodicity:
\begin{equation}
\varphi (\bm{r} + \bm{R} | \bm{0}) = \mathrm{e}^{\mathrm{i} (\pi {m_b}^2 / 2 + m_b k x)} \varphi (\bm{r} | \bm{0}).
\label{eq:TranslationGeneral}
\end{equation}
Combining Eqs.~\eqref{eq:varphi_r_r0} and \eqref{eq:TranslationGeneral}, we can obtain
\begin{equation}
\varphi (\bm{r} | \bm{r}_0 + \bm{R}) = \mathrm{e}^{\mathrm{i} (\pi {m_b}^2 / 2 + m_b k x_0)} \varphi (\bm{r} | \bm{r}_0),
\end{equation}
which shows that $\varphi (\bm{r} | \bm{r}_0)$ and $\varphi (\bm{r} | \bm{r}
_0 + \bm{R})$ are not independent; therefore, we only have to consider a set $\{ \varphi (\bm{r} | \bm{r}_0) \}_{\bm{r}_0}$ where $\bm{r}_0$ is in a unit cell of the vortex lattice. Moreover, Eqs.~\eqref{eq:varphi_r_0} and \eqref{eq:varphi_r_r0} lead to the following orthonormal relation:
\begin{equation}
\int_S \mathrm{d}^2 \bm{r} \varphi^* (\bm{r} | \bm{r}_0) \varphi (\bm{r} | \bm{r}'_0) = \delta_{\bm{r}_0, \bm{r}'_0},
\end{equation} 
where $S$ means the entire $x$-$y$ plane.

From Eqs.~\eqref{eq:varphi_r_r0} and \eqref{eq:TranslationGeneral}, we can show another relation:
\begin{equation}
\varphi (\bm{r} + \bm{R} | \bm{r}_0) = \mathrm{e}^{\mathrm{i} [\pi {m_b}^2 / 2 + m_b k x - (\bm{r}_0 \times \widehat{z}) \cdot \bm{R} / l^2]} \varphi (\bm{r} | \bm{r}_0).
\label{eq:TranslationGeneral_r0}
\end{equation}
Defining a momentum vector corresponding to $\bm{r}_0$ as
\begin{equation}
\bm{k}_0 = - \frac{\bm{r}_0 \times \widehat{z}}{l^2} \ \left( \Leftrightarrow \ \bm{r}_0 = l^2 \bm{k}_0 \times \widehat{z} \right),
\label{Definition_k0}
\end{equation}
we can rewrite Eq.~\eqref{eq:TranslationGeneral_r0} as
\begin{equation}
\varphi (\bm{r} + \bm{R} | \bm{r}_0) = \mathrm{e}^{\mathrm{i} (\pi {m_b}^2 / 2 + m_b k x + \bm{k}_0 \cdot \bm{R})} \varphi (\bm{r} | \bm{r}_0).
\end{equation}
Combination of Eqs.~\eqref{eq:TranslationGeneral} with \eqref{eq:TranslationGeneral_r0} leads to
\begin{equation}
\varphi^* (\bm{r} + \bm{R} | \bm{0}) \varphi (\bm{r} + \bm{R} | \bm{r}_0) = \mathrm{e}^{\mathrm{i} \bm{k}_0 \cdot \bm{R}} \varphi^* (\bm{r} | \bm{0}) \varphi (\bm{r} | \bm{r}_0),
\end{equation}
which means that $\varphi^* (\bm{r} | \bm{0}) \varphi (\bm{r} | \bm{r}_0)$ is a Bloch function with a lattice momentum vector $\bm{k}_0$; therefore, we can expand this function as~\cite{Ikeda_1990, *Ikeda_1992}
\begin{equation}
\varphi^* (\bm{r} | \bm{0}) \varphi (\bm{r} | \bm{r}_0) = \frac{1}{L_x L_y} \sum_{\bm{K}} \mathrm{e}^{\mathrm{i} (\bm{k}_0 + \bm{K}) \cdot \bm{r}} F_{\bm{K}} (\bm{k}_0),
\label{eq:Expansion_varphistarvarphi}
\end{equation}
where $\bm{K}$ is a reciprocal lattice vector, which can be written with a certain lattice vector $\bm{R} = m_a \bm{a} + m_b \bm{b}$, as
\begin{equation}
\bm{K} = - \frac{\bm{R} \times \widehat{z}}{l^2}.
\end{equation}

Applying the Fourier transformation to Eq.~\eqref{eq:Expansion_varphistarvarphi}, we obtain
\begin{equation}
F_{\bm{K}} (\bm{k}_0) = \int_S \mathrm{d}^2 \bm{r} \, \mathrm{e}^{- \mathrm{i} (\bm{k}_0 + \bm{K}) \cdot \bm{r}} \varphi^* (\bm{r} | \bm{0}) \varphi (\bm{r} | \bm{r}_0).
\label{eq:FK}
\end{equation}
Using the definition of $\varphi (\bm{r} | \bm{r}_0)$ [Eqs.~\eqref{eq:varphi_r_r0} and \eqref{eq:varphi_r_0}] in Eq.~\eqref{eq:FK}, we can derive the specific form of $F_{\bm{K}} (\bm{k}_0)$,
\begin{widetext}
\begin{equation}
F_{\bm{K}} (\bm{k}_0) = \exp \left\{ l^2 \left[ - \frac{(\bm{K} + \bm{k}_0)^2}{4} - \frac{\mathrm{i}}{2} \left( \frac{{K_x}^2}{\sqrt{3}} + K_x K_y + k_{0,x} k_{0,y} - (\bm{K} \times \bm{k}_0)_z \right) \right] \right\}
\label{eq:FKExplicit}
\end{equation}
\end{widetext}

Let us divide the order-parameter field $\psi (\bm{r}, z)$ (note that in this Appendix $\bm{r}$ represents a coordinate vector in the $x$-$y$ plane) into the mean-field vortex-lattice state $\varphi (\bm{r} | \bm{0}) / \sqrt{L_z}$ and the fluctuation around it:
\begin{equation}
\psi (\bm{r}, z) = \overline{a} \varphi (\bm{r} | \bm{0}) \frac{1}{\sqrt{L_z}} + \sum_{\bm{k}_0, q_z} a_{\bm{k}_0 q_z} \varphi (\bm{r} | \bm{r}_0) \frac{\mathrm{e}^{\mathrm{i} q_z z}}{\sqrt{L_z}}.
\label{eq:OPExpansionVL}
\end{equation}
Here we choose the vortex-lattice state with $\bm{r}_0 = \bm{0}$ as a spontaneously translational-symmetry broken state. Also, $\overline{a}$ and $a_{\bm{k}_0 q_z}$ represent the mean-field and fluctuation amplitudes, respectively.

The mean-field amplitude $\overline{a}$ is determined by minimizing the GL functional $\mathcal{F}_\mathrm{GL}$ [Eq.~\eqref{eq:GLFunctionalContinuous}], leading to the following expression: 
\begin{equation}
\overline{a} = \sqrt{L_x L_y L_z \frac{|\alpha_{\bm{q}_H}|}{\beta_\mathrm{A} \beta}},
\end{equation}
where we assume that $\alpha_{\bm{q}_H} < 0$, or $H < H_\mathrm{c2} (T)$, so that the mean-field approximation leads to the vortex-lattice solution. Here we choose the vortex-lattice state with $\arg (\overline{a}) = 0$ as a spontaneously $U(1)$-symmetry broken state. Here $\beta_\mathrm{A}$ is the Abrikosov factor, which characterizes the triangular lattice structure: $\beta_\mathrm{A} = \langle |\varphi (\bm{r} | \bm{0})|^4 \rangle_S / [\langle |\varphi (\bm{r} | \bm{0})|^2 \rangle_S]^2$, with a spatial average in the $x$-$y$ plane $\langle \cdots \rangle_S = (L_x L_y)^{-1} \int_S \mathrm{d}^2 \bm{r} (\cdots)$.

Using the expanded form of the order-parameter field [Eq.~\eqref{eq:OPExpansionVL}] in the GL functional [Eq.~\eqref{eq:GLFunctionalContinuous}] and diagonalizing the Gaussian-fluctuation (second-order with respect to $\{ a_{\bm{r}_0 q_z} \}$) terms, we obtain
\begin{equation}
\mathcal{F}_\mathrm{GL} = \mathcal{F}_\mathrm{GL}^\mathrm{MF} + \mathcal{F}_\mathrm{GL}^\mathrm{Gauss} + \mathcal{F}_\mathrm{GL}^\mathrm{nonGauss},
\end{equation}
where
\begin{equation}
\mathcal{F}_\mathrm{GL}^\mathrm{MF} = -L_x L_y L_z \frac{|\alpha_{\bm{q}_H}|^2}{2 \beta_\mathrm{A} \beta},
\end{equation}
\begin{equation}
\mathcal{F}_\mathrm{GL}^\mathrm{Gauss} = \sum_{\bm{k}_0, q_z > 0, m = \pm} \left( E_{\bm{k}_0}^{(m)} + \gamma {q_z}^2 \right) \left| a_{\bm{k}_0 q_z}^{(m)} \right|^2,
\end{equation}
and $\mathcal{F}_\mathrm{GL}^\mathrm{nonGauss}$ involves other terms corresponding to non-Gaussian fluctuation. In the following, we neglect the non-Gaussian fluctuation $\mathcal{F}_\mathrm{GL}^\mathrm{nonGauss}$ and concentrate on the Gaussian fluctuation $\mathcal{F}_\mathrm{GL}^\mathrm{Gauss}$. The fluctuation amplitude $a_{\bm{k}_0 q_z}^{(m)}$ is defined as
\begin{equation}
a_{\bm{k}_0 q_z}^{(\pm)} = \frac{1}{\sqrt{2}} \left( a_{\bm{k}_0 q_z} \pm a_{-\bm{k}_0, -q_z} \right),
\end{equation}
and the fluctuation energy of each mode $E_{\bm{r}_0}^{(m)}$ is obtained as
\begin{align}
E_{\bm{k}_0}^{(\pm)} = \frac{|\alpha_{\bm{q}_H}|}{\beta_\mathrm{A}} & \left[ 2 \sum_{\bm{K}} |F_{\bm{K}} (\bm{k}_0)|^2 - \sum_{\bm{K}} |F_{\bm{K}} (\bm{0})|^2 \right. \nonumber \\
& \left. \pm \left| \sum_{\bm{K}} F_{\bm{K}} (\bm{k}_0)^2 \right| \right],
\end{align}
where $F_{\bm{K}} (\bm{k}_0)$ is given in Eq.~\eqref{eq:FKExplicit}. We can show that $F_{\bm{K}} (\bm{0}) \in \mathbb{R}$, so that $E_{\bm{0}}^{(-)} = 0$, which shows that the fluctuation mode represented as $a_{\bm{k}_0 q_z}^{(-)}$ is massless (corresponding to the incompressible shear mode of the vortex lattice~\cite{Moore_1989, Ikeda_1990}). Since the massless mode is expected to be dominant in considering the melting transition~\cite{Moore_1989}, we take into account the contribution of the massless mode $a_{\bm{k}_0 q_z}^{(-)}$ and neglect that of the massive mode $a_{\bm{k}_0 q_z}^{(+)}$. Moreover, to consider the long-wavelength and low-energy contribution of the massless mode, we expand the fluctuation energy $E_{\bm{k}_0}^{(-)}$ with respect to $\bm{k}_0$:
\begin{equation}
E_{\bm{k}_0}^{(-)} = \frac{\gamma_\mathrm{A} |\alpha_{\bm{q}_H}|}{\beta_\mathrm{A}} l^4 {k_0}^4 + \mathcal{O} ({k_0}^6).
\label{eq:EMinus}
\end{equation}
Here $\gamma_\mathrm{A}$ is a numerical factor related to the triangular-lattice structure:
\begin{align}
\gamma_\mathrm{A} & = \sum_{\bm{K}} \mathrm{e}^{- l^2 K^2 / 2} \left\{ \frac{1}{12} \left[ \frac{3}{8} l^4 K^4 - 3 l^2 K^2 + 3 \right] - \frac{1}{8} \right\} \nonumber \\
& \simeq 0.119.
\end{align}
To derive Eq.~\eqref{eq:EMinus}, we use the following properties with an arbitrary function $f (K) = f (|\bm{K}|)$ due to a six-fold rotational symmetry of the reciprocal lattice space:
\begin{equation}
\left\{
\begin{aligned}
\sum_{\bm{K}} (\bm{K} \cdot \bm{k}_0)^2 f(K) & = \sum_{\bm{K}} \frac{1}{2} K^2 {k_0}^2 \\
\sum_{\bm{K}} (\bm{K} \cdot \bm{k}_0)^4 f(K) & = \sum_{\bm{K}} \frac{3}{8} K^4 {k_0}^4.
\end{aligned}
\right.
\end{equation}
In the following, therefore, we focus on the following functional:
\begin{equation}
\mathcal{F}_\mathrm{GL}^{\mathrm{Gauss}(-)} = \sum_{\bm{k}_0, q_z > 0} \left( \frac{\gamma_\mathrm{A} |\alpha_{\bm{q}_H}|}{\beta_\mathrm{A}} l^4 {k_0}^4 + \gamma {q_z}^2 \right) \left| a_{\bm{k}_0 q_z}^{(-)} \right|^2.
\label{eq:GLFunctionalLowEnergy}
\end{equation}
It has been proved~\cite{Ikeda_1990, *Ikeda_1992} that this form of the dispersion relation of the massless mode of the vortex lattice in type-II limit remains valid when the higher Landau-level modes ($N \geq 1$) are included.

Since the relative fluctuation $2^{-1/2} a_{\bm{k}_0 q_z}^{(-)} / |\overline{a}|$ can be regarded as an angular change of the vortex lattice $\theta_{\bm{k}_0 q_z}$~\cite{Moore_1989}, we can rewrite $\mathcal{F}_\mathrm{GL}^{\mathrm{Gauss}(-)}$ [Eq.~\eqref{eq:GLFunctionalLowEnergy}] as
\begin{align}
\mathcal{F}_\mathrm{GL}^{\mathrm{Gauss}(-)} & = L_x L_y L_z \sum_{\bm{k}_0, q_z > 0} \left( c_{66} l^4 {k_0}^4 + \rho_\mathrm{s} {q_z}^2 \right) \left| \theta_{\bm{k}_0 q_z} \right|^2 \nonumber \\
& = \frac{1}{2} \int_S \mathrm{d}^2 \bm{r} \int_{0}^{L_z} \mathrm{d} z \, \left[ c_{66} l^4 ({\nabla_{\perp}}^2 \theta)^2 + \rho_\mathrm{s} (\partial_z \theta)^2 \right].
\label{eq:GLFunctionalFinal}
\end{align}
Here, $\theta (\bm{r}, z) = \sum_{\bm{k}_0, q_z} \mathrm{e}^{\mathrm{i} (\bm{k}_0 \cdot \bm{r} + q_z z)} \theta_{\bm{k}_0 q_z}$ is a real-space phase field related to the vortex-lattice displacement field $\bm{u} (\bm{r}, z)$~\cite{Moore_1989} as
\begin{equation}
\left\{
\begin{aligned}
u_x & = l^2 \partial_y \theta \\
u_y & = - l^2 \partial_x \theta.
\end{aligned}
\right.
\end{equation}
This relation indicates that the vortex-lattice deformation corresponding to the massless mode $a_{\bm{k}_0 q_z}^{(-)}$ represents an incompressible shear mode: $\nabla_{\perp} \cdot \bm{u} (\bm{r}, z) = 0$~\cite{Moore_1989}.
Also, $c_{66}$ and $\rho_\mathrm{s}$ represent the shear modulus of the vortex lattice and the superfluid density defined as the response quantity in the $z$ direction, respectively:
\begin{equation}
c_{66} = \frac{2 \gamma_\mathrm{A} |\alpha_{\bm{q}_H}|^2}{{\beta_\mathrm{A}}^2 \beta},
\end{equation}
and
\begin{equation}
\rho_\mathrm{s} = \frac{2 |\alpha_{\bm{q}_H}| \gamma}{\beta_\mathrm{A} \beta}.
\end{equation}

The mean square displacement of the vortex lattice $d^2 = \langle |\bm{u} (\bm{r})|^2 \rangle$ is calculated as
\begin{equation}
d^2 = \langle |\bm{u}|^2 \rangle = l^4 \left< (\nabla_{\perp} \theta)^2 \right> = 2 l^4\sum_{\bm{k}_0, q_z > 0} {k_0}^2 \langle |\theta_{\bm{k}_0 q_z}|^2 \rangle.
\end{equation}
Here, $\langle \cdots \rangle$ means the ensemble average with respect to the low-energy GL functional $\mathcal{F}_\mathrm{GL}^{\mathrm{Gauss}(-)}$ [Eq.~\eqref{eq:GLFunctionalFinal}], and thus we can obtain the following formula:
\begin{equation}
d^2 = \frac{l^4}{L_x L_y L_z} \sum_{\bm{k}_0, q_z} \frac{T {k_0}^2}{c_{66} l^4 {k_0}^4 + \rho_\mathrm{s} {q_z}^2}.
\end{equation}
Since the summation about $q_z$ is convergent, we take ${L_z}^{-1} \sum_{q_z} (\cdots) \rightarrow (2 \pi)^{-1} \int_{- \infty}^{\infty} \mathrm{d} q_z (\cdots)$. On the other hand, since the summation about $\bm{k}_0$ is not convergent if $k_0 \rightarrow \infty$, we simply replace the summation with an integration over an area corresponding to the first Brillouin zone: $(L_x L_y)^{-1} \sum_{\bm{k}_0} (\cdots) \rightarrow (2 \pi)^{-1} \int_{0}^{\sqrt{2} / l} \mathrm{d} k_0 \, k_0 (\cdots)$. These replacements lead to the following simple expression:
\begin{equation}
d^2 = \frac{T}{4 \pi \sqrt{\rho_\mathrm{s} c_{66}}}.
\end{equation}
Using the Lindemann criterion~\cite{Moore_1989}, we can expect that the vortex lattice can melt into the vortex liquid when a condition $d = c \times l$ is satisfied [note that the magnetic length $l$ corresponds to the unit-cell size (see Fig.~\ref{fig:UnitCell})], where $c = \mathcal{O} (0.1)$ is a phenomenological parameter. Introducing a dimensionless magnetic field $h = 2 \pi \mu_0 H / \phi_0 = l^{-2}$ (note that the lattice constant is set to unity), we obtain the equation [Eq.~\eqref{eq:VortexLatticeFormastionField}] describing the melting-transition field, or the vortex-lattice-formation field, $H_\mathrm{melt}$:
\begin{equation}
\frac{T}{4 \pi \sqrt{\rho_\mathrm{s} c_{66}}} = \frac{c^2}{h}.
\end{equation}




\input{PaperMelting.bbl}


\end{document}

%% file: PaperMelting.bbl
%